\documentclass[useAMS,usenatbib,referee]{mnras}
\usepackage{graphicx}
\usepackage{booktabs}
\usepackage{longtable}
\usepackage{rotating}
\usepackage{lscape}
\usepackage{subfigure}
\usepackage{subfloat}

\begin{document}
\bibliographystyle{plainnat}

\title[Infrared Study of PAGB Stars]
{An Infrared Photometric and Spectroscopic Study of Post-AGB Stars}
\author[V. Venkata Raman et al]{V. Venkata
Raman${^1}$\thanks{vvenkat@prl.res.in}, B.G. Anandarao${^1}$
\thanks{anand@prl.res.in}, P. Janardhan${^1}$\thanks{jerry@prl.res.in} 
and R. Pandey${^2}$\thanks{pandey.rj@gmail.com} \\
$^1$Physical Research Laboratory (PRL),
Navrangpura, Ahmedabad - 380009, India \\
$^2$Physics Department, M.L.S. University, Udaipur, India}

\date{}

\pagerange{\pageref{firstpage}--\pageref{lastpage}} \pubyear{2017}

\maketitle

\label{firstpage}

\begin{abstract}
We present here $Spitzer$ mid-infrared (IR) spectra and modeling of the spectral energy distribution (SED) of a selection of 
post-Asymptotic Giant Branch (PAGB) stars.
The mid-IR spectra of majority of these sources showed spectral features such as polycyclic aromatic hydrocarbons (PAHs) and silicates in emission. 
Our results from SED modeling showed interesting trends of dependency between the photospheric and circumstellar parameters.
A trend of dependency is also noticed between the ratios of equivalent widths (EWs) of various vibrational modes of PAHs  
and the photospheric temperature T$_{*}$ and model-derived stellar parameters for the sample stars.
The PAGB mass loss rates derived from the SED models are found to be higher than those for the AGB stars.
In a few objects, low and high excitation fine structure emission lines were identified, indicating their advanced stage of
evolution. Further, IR vibration modes of fullerene (C$_{60}$) were detected for the first time in the PAGB star IRAS 21546+4721.

\end{abstract}


\section{Introduction}
Post-Asymptotic Giant Branch (PAGB) stars or proto-Planetary Nebulae (PPNe) represent the short-lived,
rapidly varying and least understood phase
during the evolution of intermediate mass stars towards planetary nebulae (PNe) stage
(\citet{Vanwinckel03}; \citet{Kwok07} and references therein). The evolutionary phases of PAGB and PPNe are 
nearly indistinguishable and here onwards
we treat the two to be the same.
These sources exhibit infrared excess and molecular line emission from dust
and molecules in the circumstellar envelopes formed by the massloss during the AGB phase.
Circumstellar dust in PAGB stars cause 
significant extinction in the visible region.  
The exact composition of the circumstellar dust grains is still a matter of debate (e.g., see 
\citet{Cerrigone09} and \citet{Zhang10}).
The evolution during the PAGB phase can be 
identified by several spectroscopic signatures - from molecular hydrogen emission to atomic hydrogen recombination 
lines as well as forbidden transitions from several atoms and ions. Further, the onset of UV radiation can also be 
inferred from the detection of circumstellar Polycyclic Aromatic Hydrocarbons (PAHs) that are excited by absorption of
soft UV radiation in carbon rich PAGB stars (e.g., \citet{Kwok04}, \citet{Tielens05}). 
Investigation of circumstellar matter and their chemical synthesis can 
therefore help in probing and understanding the early stages of PN formation.

Pioneering studies had been made and reported by \citet{Kwok99}, \citet{Hrivnak00} and the references therein
on the infrared spectra of PAGB stars with particular emphasis on 
PAH features and other as yet unidentified features as well as 
on molecular hydrogen lines (e.g., \citet{Kelly05} and the references therein). 
In the present study, we model the SEDs for a fairly large sample of PAGB stars  
using the available archival data covering the visible to far-infrared regions. 
We then endeavour to find if the model-derived
circumstellar parameters indicate possible dependency on evolution of the objects. 

The sample selection of the PAGB stars used in the present study is described in Section 2. 
In Section 3, we present mid-IR spectra of the sample sources. In Section 4 
we present modeling of SEDs and in Section 5 we 
discuss the possible correlation between the circumstellar and photospheric parameters 
obtained from the SED models and also between the various PAH modes of vibration with the evolution of the central star.
In Section 6 we identify a few transition objects in our sample through the detection of 
low and high excitation fine structure lines in the mid-IR spectra. 
In Section 7 we present the detection of fullerenes in one of the sample PAGB star.
In Section 8 we summarise the conclusions of the present work. 

\section{Sample of PAGB stars and photometric archival data:} 

\begin{figure}
\centering
\includegraphics[width = 5.0in]{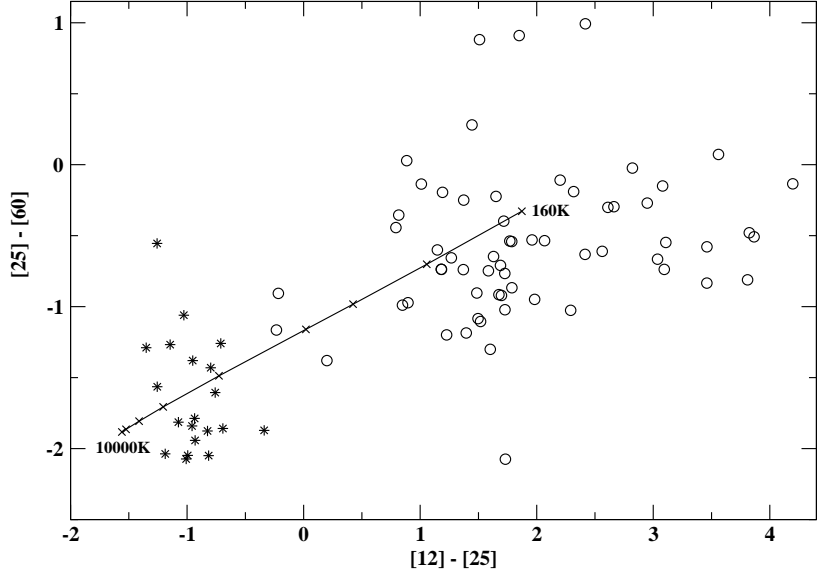}
   \caption[IRAS color-color diagram of the sample stars]{IRAS color-color diagram for the stars in our sample. 
   PAGB stars (open circles) occupy positions that have
    very cool circumstellar shells. The AGB stars (asterisks) from \citet{Venkataraman08} are also shown for comparison. 
    The crosses along the line represent the black-body temperatures of 10000, 5000, 2000, 1000, 500, 300, 250, 200, 160K
     (see \citet{Vanderveen88}) in the color-color space.}
\label{irasccd}%
\end{figure}

The sample of PAGB stars studied were selected 
from \citet{Szczerba07}. Based on their $JHK$-band magnitudes 
being brighter than 13-14, a total of about 71 PAGB 
stars were identified for the present study.
From the classification scheme based on the IRAS [25-60] vs [12-25] colour-colour diagram (shown in Fig \ref{irasccd}, 
see \citet{Vanderveen88}), 
it is clear that the objects selected for the present study 
are much more evolved than the sources in their AGB phase considered
in our earlier work (\citet{Venkataraman08}).
Spectral types for a number of PAGB candidates were taken from SIMBAD. 
For some of the objects without spectral type information from SIMBAD, we have used the spectral classification by the 
optical survey of \citet{Suarez06}. Some of the 
objects in the selected PAGB star samples are identified as "transition objects" by \citet{Suarez06}. 
It may be noted that there exist small (and in some cases, large) differences in spectral types 
of some of the common objects between SIMBAD and \citet{Suarez06}.

The photometric data required for constructing the SEDs in the 
entire visible to far-infrared region were obtained from several archives. 
The data on visible $UBVRI$ photometry were obtained from VIZIER archives; 
the near-IR $JHK$ bands from the 2MASS archive (\citet{Skrutskie06}); mid-infrared from MSX bands at 
8.3, 12.1, 14.7 and 21.3 $\mu$m (\citet{Egan03});
the 3.6, 4.5, 5.8 and 8 $\mu$m bands from IRAC-$SPITZER$ (\citet{Benjamin03}); the
mid-infrared bands at 9 and 18 $\mu$m and far-infrared bands centered around 65, 90, 140 and 160 $\mu$m were obtained
from the extensive all-sky survey from $AKARI$ (\citet{Ishihara10}); 
and finally the 12, 25, 60 and 100 $\mu$m band data were obtained from $IRAS$ (\citet{Helou88}) archives. 

\section{${Spitzer}$ spectra of PAGB objects}

$Spitzer$ public archival spectroscopic data from
Infra-Red Spectrograph (IRS) on-board the $Spitzer$ Space Telescope covering the mid-IR region
between 5.2 and 38 $\mu$m (\citet{Houck04}) were used for most of the sample stars in our present study.
Most of the data used here were from the short-low (SL) and long-low (LL) modules covering 5.13 to 14.29 $\mu$m
and 13.9 to 39.9 $\mu$m respectively. For a few sources, the short-high (SH - 9.89 to 19.51 $\mu$m) and long-high 
(LH - 18.83 to 37.14 $\mu$m) modules were also used.
\begin{subfigures}
	\begin{figure}[]{
	\includegraphics[width = 7.4in]{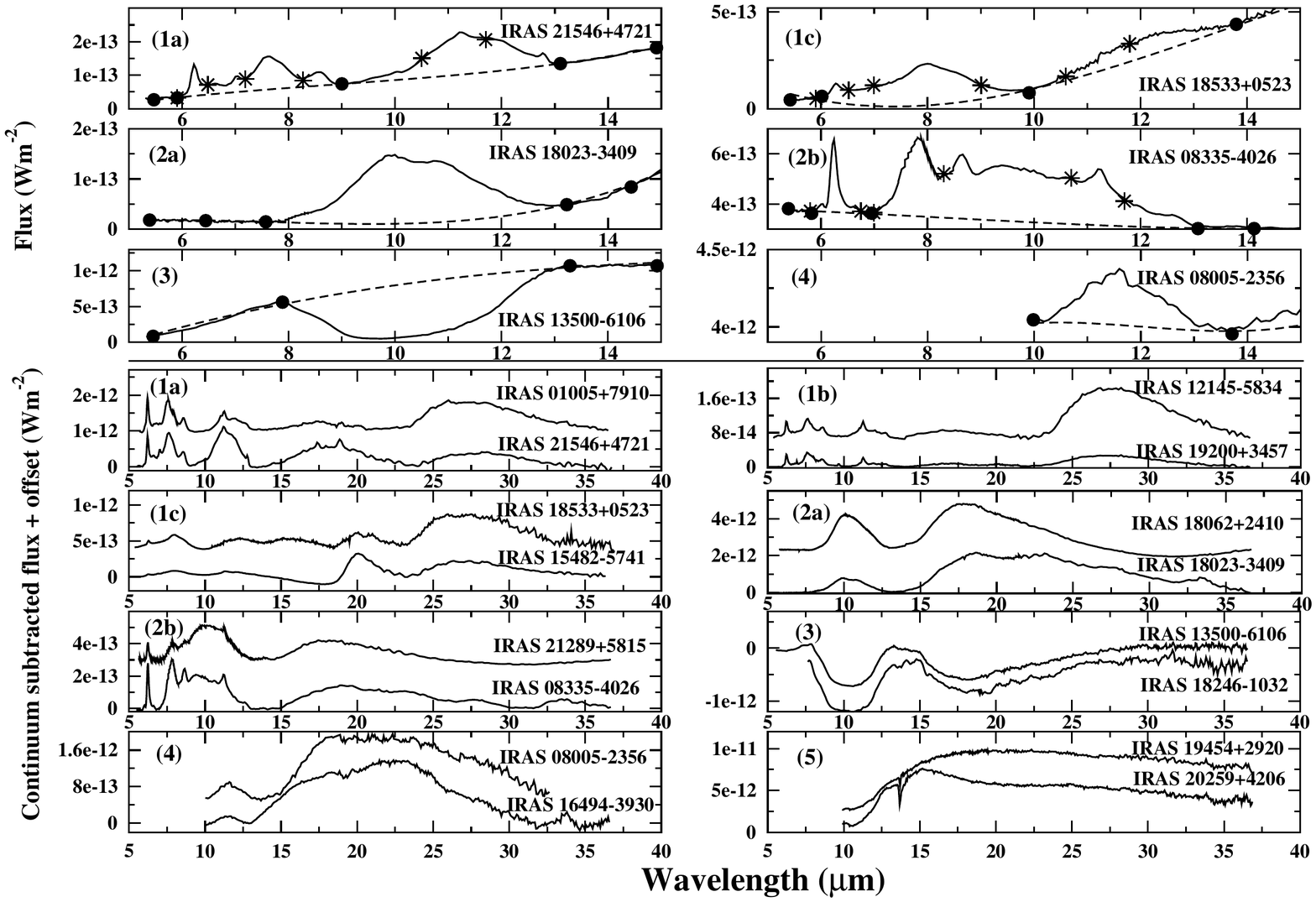}}
	\vspace{-10mm}
	\caption{\label{first}Extracted raw spectra (solid curves) 
	    with pivotal points shown in filled circles and asterisks for defining the global and plateau continuum  
	    are shown in the top three (left and right) panels. Plotted in the bottom eight  panels (left and right) are the continuum-subtracted $Spitzer$ IRS spectra of the PAGB stars
	   showing (1a) strong PAH features, (1b) weak PAH features (1c) blended PAH feature,
	   (2a) silicate dust emission features, (2b) PAH + silicate emission (mixed chemistry), 
	   (3) silicate absorption (4) prominent broad bump around 11 $\mu$m and 
	   (5) featureless spectra (continuum not subtracted)}
	\label{pagb_fig2a}
	\end{figure} 
	\begin{figure}
	\centering
	\includegraphics[width = 6.0in]{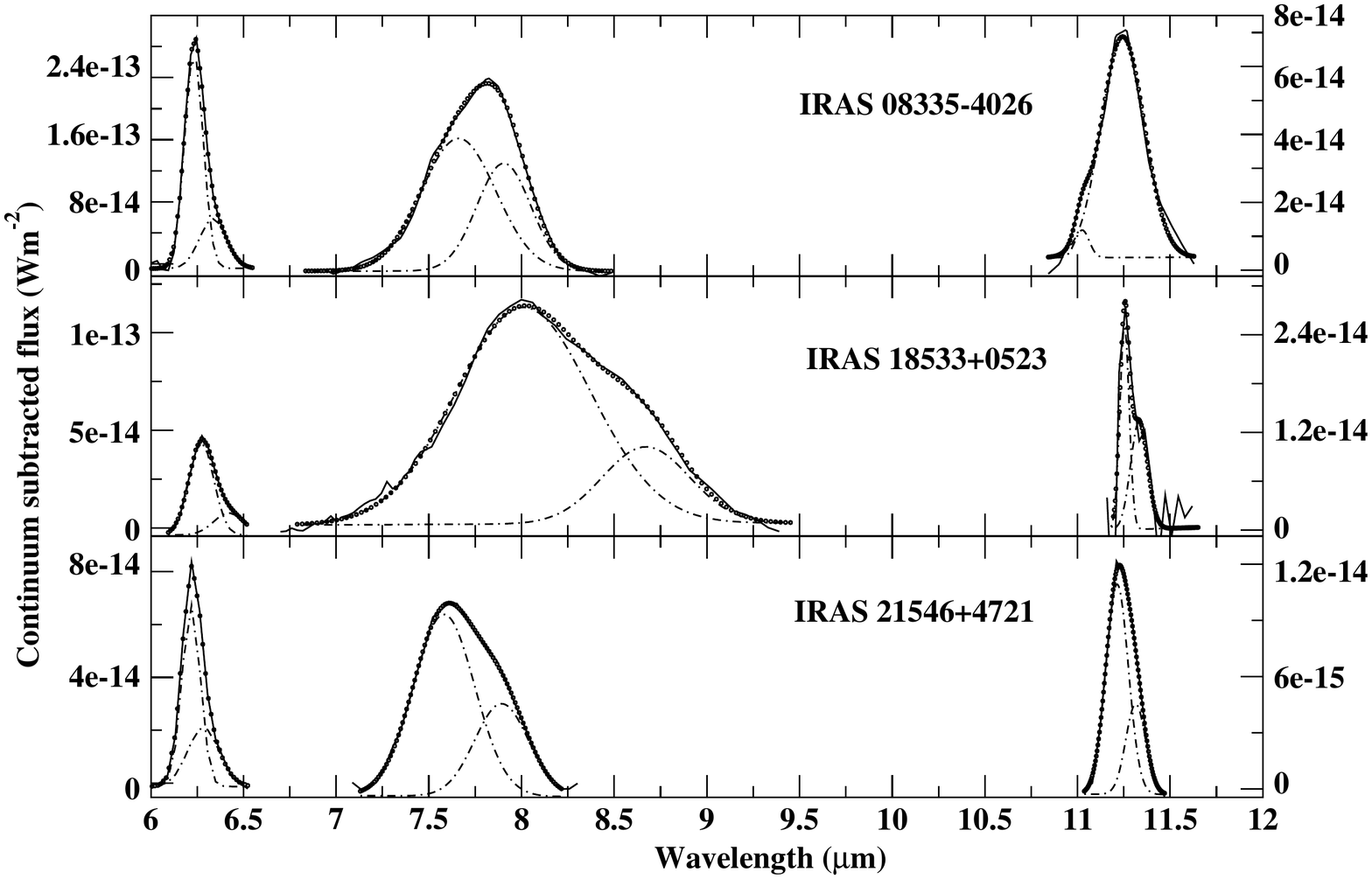}
	\caption{\label{second}Continuum-subtracted spectra of three sources chosen (from Classes 1a, 1c and 2b in Fig 2a) to illustrate multi-Gaussian decomposition of the main PAH features at 6.2, 7.7 and 11.2 $\mu$m. The fitted Gaussian profiles and their sum
		   are represented by dot-dashed and dotted lines respectively; while the solid curves represent the continuum subtracted spectra. The flux scale on the right is for 11.2 $\mu$m feature.}
	\label{pagb_fig2b}
	\end{figure}
	\end{subfigures}
The basic calibrated data (BCD) and post-BCD data from $Spitzer$ 
archive were analysed using Spectroscopic Modeling Analysis and Reduction Tool (SMART)
and $Spitzer$ IRS Custom Extraction (SPICE) software packages respectively.
The highly processed data products (HPDP) from the $ISO$ archive (\citet{Sloan03}) 
were used for a few sources in our list for which $Spitzer$ data 
was not available.  

Based on identified mid-IR spectral features, we classified our sample PAGB stars as those having
(i) strong/weak or blended PAH features  
(ii) silicate emission, in some cases along with PAH features  (iii) silicate absorption 
(iv) prominent broad bump around 11 $\mu$m with no PAH features and (v) nearly featureless spectra.
The blended PAH refers to the blend of two or more neighboring PAH lines  
to form a single broad feature.
Similar classification based on the presence of PAH (C-rich), silicate (O-rich) and a combination of 
PAH and silicate features was done by \citet{Cerrigone09}.
In addition to Cerrigone et al's classification, the present study
identifies objects that have silicate absorption and those with 11 $\mu$m bump
and thirdly those with featureless spectra.
Table \ref{pagb1} provides the list of all the programme objects 
along with their spectral data sources and the type of dust from our classification.
Figure \ref{pagb_fig2a} gives typical spectra of the PAGB candidates depicting
the above classification. The figure shows in the top three panels the raw spectra of the PAGB candidates 
on which pivotal points are marked (filled circle; see \citet{Spoon07}) to define the global continuum.
The continuum as defined by the pivotal points
on the spectra were then fitted by a second or third order polynomial for the 
entire region.   
In the figure the continuum-subtracted spectra are shown in the bottom eight panels.   
The spectra for all the individual sources in each class are given in the Appendix A.
Out of the 71 PAGB sources considered in the present study, 11 objects do not fall under any of the classes.
In addition, 2 PAGB sources showing fine structure lines have very peculiar dust features 
that do not fall in the above classification. 
Among the remaining 58 sources, about 29\% of the PAGB sources showed strong or weak/blended PAH features,
26\% showed silicate emission, 17 \% of sources showed PAH along with silicate in emission (mixed chemistry),
12\% of sources showed silicate absorption,
11\% of sources showed 11 $\mu$m feature, 5\% of sources showed featureless spectra. 

Some details of the sources in different classes or categories are described below 
with reference to the spectra beyond 20 $\mu$m, up to 38 $\mu$m. 
In this spectral region the most prominent features seen are at 21 and 30 $\mu$m.
Among the 17 sources that fall under the PAH category, 9 show both 21 \& 30 $\mu$m features,
3 sources show only 30 $\mu$m feature and 1 source shows only 21 $\mu$m feature.
The sources IRAS 04296+3429, IRAS 05113+1347, IRAS 06530-0213, IRAS 14429-4539, IRAS 15482-5741, IRAS 18533+0523, IRAS 19200+3457, IRAS 22223+4327
and IRAS 23304+6147 show both 21 \& 30 $\mu$m features. The sources IRAS 01005+7910, IRAS 12145-5834 and IRAS 21546+4721
show only 30 $\mu$m feature while IRAS 07134+1005 show only 21 $\mu$m feature. 
All these sources show strong, blended or weak PAH features also.
The presence of 21 $\mu$m along with the observed PAH features thus depicts their carbon rich chemistry.

Silicate emissions at both 9.7 \& 18 $\mu$m were observed in 15 sources. 
The peak position of silicate emission occurs usually at 9.7 $\mu$m, but we found that 
in some sources the peak occurs at longer wavelengths ($\approx$ 10.1 to 10.2 $\mu$m). 
\citet{Bouwman01} suggested an increase in grain size 
responsible for the change in the 9.7 $\mu$m peak position towards longer wavelengths. 
In the objects that are categorized under this section a number of sources showed features at 27 and 33 $\mu$m attributed to 
crystalline olivines (see \citet{Jaeger98}).
These sources do not show the 21 and 30 $\mu$m features. 
Silicate emission along with PAH (mixed chemistry) has been observed in 10 sources.
A number of sources in this category also showed the presence of crystalline olivines at 27 and 33 $\mu$m.
Waters et al (1998) proposed a scenario for mixed chemistry in which a presently carbon-rich star 
may retain in its outer layers the circumstellar oxygen-rich gas due to an earlier phase of mass loss. 
This scenario can then account for both carbon-rich (PAH emission) and oxygen-rich (silicate emission) signatures.
Using SOFIA infrared imager observations combined with 3D photoionization and dust radiative transfer modeling, 
\citet{Guzman15} also  
 demonstrated that oxygen-rich (silicate) dust occurs in the outer regions of the planetary nebula BD +30$^{\circ}$ 3639 while the
 carbon-rich material (PAHs) is located in the inner parts.
 These earlier studies suggest the emergence of mixed chemistry possibly caused by dredge-up of carbon from inner regions 
 of the central star towards the end of the AGB stage. 

The absorption due to silicates at 9.7 and 18 $\mu$m has been observed in 7 sources. The spectral types of all these sources
are not known due to the possibility of their photospheres being heavily enshrouded by circumstellar shells. 
\citet{Garcia03} suggested an increased thickness of the circumstellar
shell responsible for the absorption features in oxygen-rich sources. 

The broad 11 $\mu$m feature usually attributed to SiC has been observed in IRAS 08005-2356, IRAS 11339-6004,
IRAS 14325-6428, IRAS 14341-6211, IRAS 16494-3930 and IRAS 17359-2902. It may be noted that these sources
do not show PAH emissions. It is generally believed that SiC emission features
are observed in sources with very thick dust shells formed from carbon-based dust grains. 
\citet{Garcia03} attributed the observed increase in the 
strength of the SiC feature in carbon stars to the increased mass-loss rate during the AGB phase of the stars. 
\citet{Speck05} argued
that the change in the appearance of the 11 $\mu$m feature is a result of self-absorption from the thick dust shells.
In the longer wavelength region, among the six sources in this category, 
two sources show 30 $\mu$m feature; and the two other sources do not show 30 $\mu$m feature but 
show crystalline olivines. 

Among the 3 objects that are categorised as showing featureless spectra, the sources IRAS 19454+2920 \&
IRAS 20259+4206 show narrow absorption line of the carbon based molecule, C$_2$H$_2$ at 13.7 $\mu$m suggesting
their advanced stage of evolution towards a carbon-rich star; in addition to the featureless  thermal continuum due to dust. 
On the basis of the observed featureless thermal emission from the circumstellar envelope
and a weak Rayleigh-Jeans tail at the shorter wavelengths due to the central star, it may be conjectured
that objects displaying such spectral characteristics may be relatively young 
PAGB stars evolving beyond the heavily enshrouded OH/IR stars
(\citet{Waelkens03}).

The carrier candidates for the 21 and 30 $\mu$m feature are still being debated. The candidates for the 21 $\mu$m
feature include TiC (\citet{Vonhelden00}), doped SiC dust (\citet{Speck04}),
FeO (\citet{Posch04}) and PAH molecules (\citet{Justtanont96}). 
\citet{Goebel85} suggested
solid magnesium sulphide (MgS) as the possible carrier for the 30 $\mu$m feature, while \citet{Duley00} suggested 
carbon-based linear molecules with specific side groups.

For the purpose of analysing the correlations, equivalent widths (EWs) of the most prominent 
mid-infrared PAH features, namely, those at 6.2, 7.7 \& 11.2 $\mu$m, were derived 
using the Spectroscopic Modeling Analysis and Reduction Tool (SMART). 
Table \ref{pagb2} lists the EWs for all objects that showed PAH features. 
The table also lists the peak positions of the three features for each object. 
EWs of the three PAH features considered here were determined 
using the procedures adopted by \citet{Hony01}, \citet{Peeters02}
and \citet{Peeters17}.
In addition to the global continuum as shown in Fig \ref{pagb_fig2a}, the local continuum around the 
5-12 $\mu$m region is  enhanced by broad plateaus 
around 5-7 $\mu$m, 7-8 $\mu$m (8 $\mu$m bump) and 10-12 $\mu$m (see \citet{Peeters17} and the references therein). 
These plateaus are typically shown by asterisks in Fig \ref{pagb_fig2a} (see class 1a, 1c, 2b in top three panels).
The pivotal points defining the plateau continua were fitted by second or third order polynomial and
subtracted from the raw spectra. 
The resulting continuum-subtracted spectra around 6.2, 7.7 and 11.2 $\mu$m were fitted suitably by 
single/multi Gaussian profiles (using LINE FIT task in SMART) to account for the PAH emission features. 
In most cases, spectral decomposition using two Gaussian was required to account for
blended components of PAH emission at 7 and 11 $\mu$m regions, while
the 6.2 $\mu$m feature required single Gaussian in 6 objects, two in 12 and three Gaussians in 3 objects.
Figure \ref{pagb_fig2b} shows continuum-subtracted spectra of 3 objects (of Classes 1a, 1c and 2b) for which Gaussian decomposition is illustrated.
In the literature several methods had been used to extract the
intensity or EWs of the PAH features (e.g. Drude profiles (\citet{Smith07}), Lorentzian (\citet{Galliano08})
and Gaussian (\citet{Uchida00}, \citet{Hony01}, \citet{Peeters02}, \citet{Blasberger17}
and \citet{Peeters17})). In a recent study
\citet{Peeters17} concluded that for a large sample of objects
PAH correlations are independent of the method adopted. 
 The earlier studies suggested that the line fluxes were only marginally affected by the choice of the profile while they depend more on
 the choice of the continuum. Further, the line centre determination does not critically 
 depend upon the choice of profile (for e.g. \citet{Blasberger17}).
The errors in the EW of the PAH features (see Table \ref{pagb2}) were estimated using the relation, 

\begin{equation}
\sigma ^2_T(W)=M.\left ( \frac{h_\lambda }{S/N} \right )^2.\frac{\tilde{F}_j}{\tilde{F}_c}+\left [ \frac{\sigma (\tilde{F}_c)}{\tilde{F}_c} .(\Delta\lambda - W)\right ]^2
\end{equation}
\\
from \citet{Chalabaev83}, where 
h$_{\lambda}$ is the spectral resolution per pixel, M is the number of pixels, S/N the
signal to noise ratio,
$\frac{\tilde{F}_j}{\tilde{F}_c}$ represent the ratio of the measured flux over the continuum
summed over the number of pixels  
and $\sigma(\tilde{F}_c)$ 
represent the uncertainty in the continuum, $\Delta\lambda$=$\lambda_{1}$-$\lambda_{2}$ is 
the wavelength range covered
by the feature and W represents the EW. 

\section{Modeling of SEDs:} 

 The SEDs of the sample stars constructed from archival photometric data, were modeled using 
the software DUSTY (\citet{Ivezic99}).  
The input parameters for each object consist of stellar photospheric temperature (T$_{*}$), temperature of the 
inner region of the dust shell (T$_{d}$), composition of the dust grains, density distribution and the optical depth at 0.55 $\mu$m ($\tau_{0.55}$). The composition of the dust considered here for the individual 
sources is based on the observed spectral features in their $Spitzer$ IRS spectra.
The parameters that are fixed are T$_{*}$, dust particle size limits and size distribution index; and density distribution.
The variable parameters are T$_{d}$, composition of dust grains (for continuum we mainly used either graphite (Gr) or 
amorphous carbon (AmC)), optical depth at 0.55 $\mu$m ($\tau_{0.55}$).
The code solves the radiative transfer problem for a source
embedded in a spherically symmetric dusty envelope. The important output parameters of the code are
the total flux as a function of wavelength, the inner shell radius r$_{1}$, the ratio of inner shell radius and the 
radius of central source, r$_{1}$/r$_{c}$, terminal velocity and upper limit of the mass of the source. The model also gives
an estimate of the total mass loss rate (dM$_{gas+dust}$/dt=\.{M}) for envelope expansion caused by radiatively driven 
winds with an uncertainity of 30 \%.
\citet{Meixner97} and \citet{Mishra16} used axisymmetric models
to infer mass-loss rates. We find a good agreement with these models for some sources that are common
in our sample.
 
The model fluxes are compared with the observed values,  
by normalising with the flux at either 2 $\mu$m (K band) or at 
9 $\mu$m band (of MSX or $AKARI$ or IRAC) depending upon the distribution of observed data points in the infrared range.
T$_{*}$ was taken from \citet{Cox00} and \citet{Lang99}
for objects whose spectral types are identified. For all the other objects (25 in number) whose spectral types are not available, 
we have estimated T$_{*}$ 
from model best fits. 
For a majority of programme objects, the best fit degeneracy is very minimal and 
can be gauged by the uncertainty on
T$_{d}$ ($<$ 5 \%) and  $\tau_{0.55}$ (about 10 \%), grain composition that contributes for the continuum ($<$ 10 \%).  
 A criterion for the best fit was chosen as,  
$\chi^{2}$ = $\Sigma_{i}$[(F$_{Oi}$ - F$_{Mi})^{2}$/F$_{Mi}]$ $\leq$ 10$^{-12}$, where $\Sigma_{i}$ indicates 
summing over all the data points ($i$=1 to n, n being the number of data points for a particular source)
and F$_{Oi}$ and F$_{Mi}$ are the observed and model fluxes at i$^{th}$ data point respectively.
For each object, several models were generated by tweaking the
variable parameters (T$_{d}$, grain composition and $\tau_{0.55}$) and we chose the best model as
determined by the chi-square criterion as well as by visual inspection. 
Barring a few, for most sources in our sample, 
the limit on $\chi^{2}$ had resulted in (visually) reasonably good fits, except for a few sources.  
It may be noted that the spectral class differences in some of the objects that are common in SIMBAD and 
\citet{Suarez06} did not alter the model-derived parameters by more than 10\%.
 \begin{figure}
\centering
\includegraphics[scale=0.65]{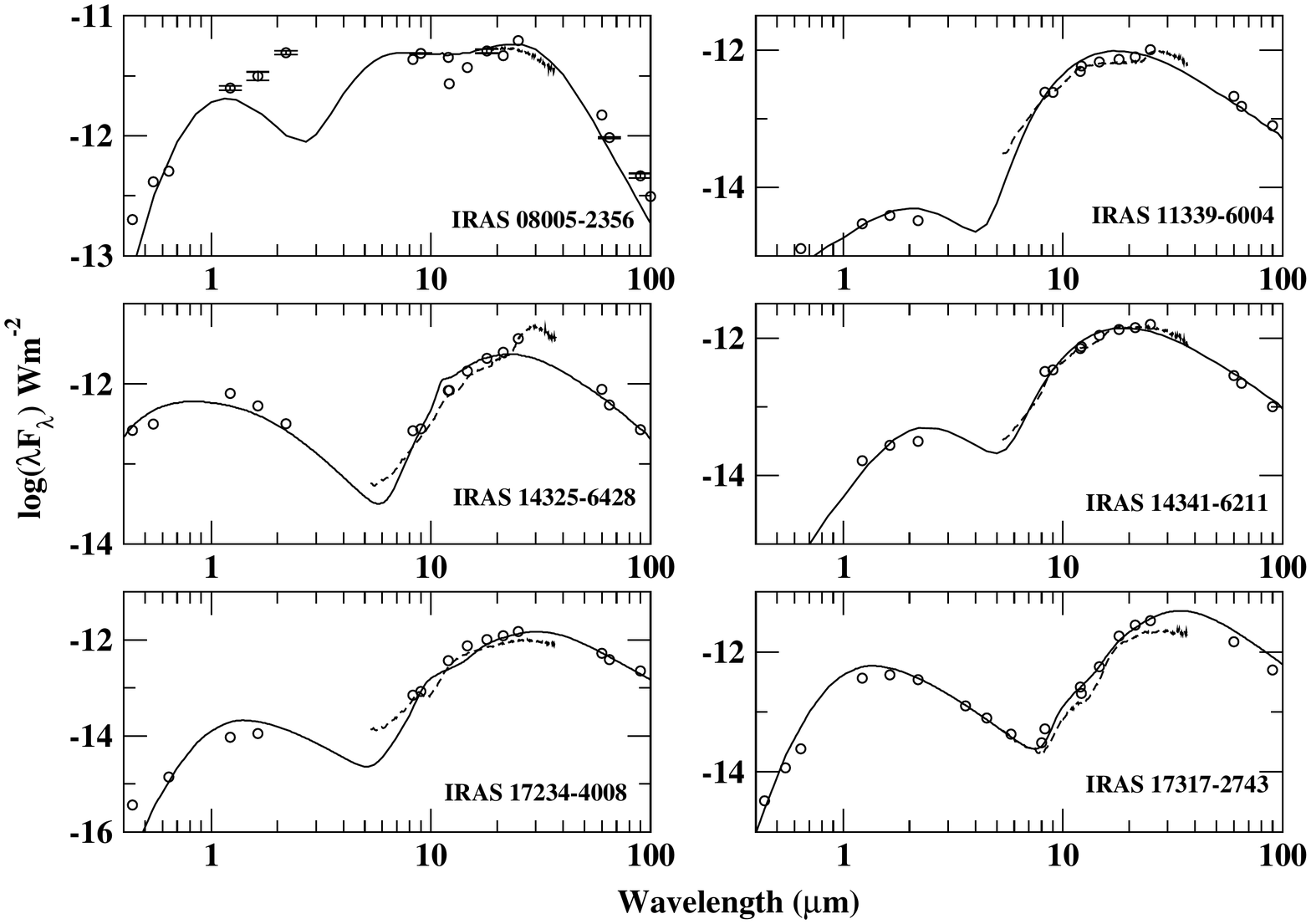}
   \caption[Typical DUSTY models of PAGB with single spherical shells]{Model SEDs with single spherical shells using DUSTY (solid line) for a few PAGB sources compared with observed data from literature (open circles). The ${Spitzer}$ IRS spectra (dashed line) are shown for comparison. Being
   in the log scale the error bars are of the point size.}
\label{ppne_dusty_single}%
\end{figure}
\begin{figure}

\begin{center}
\includegraphics[scale=0.65]{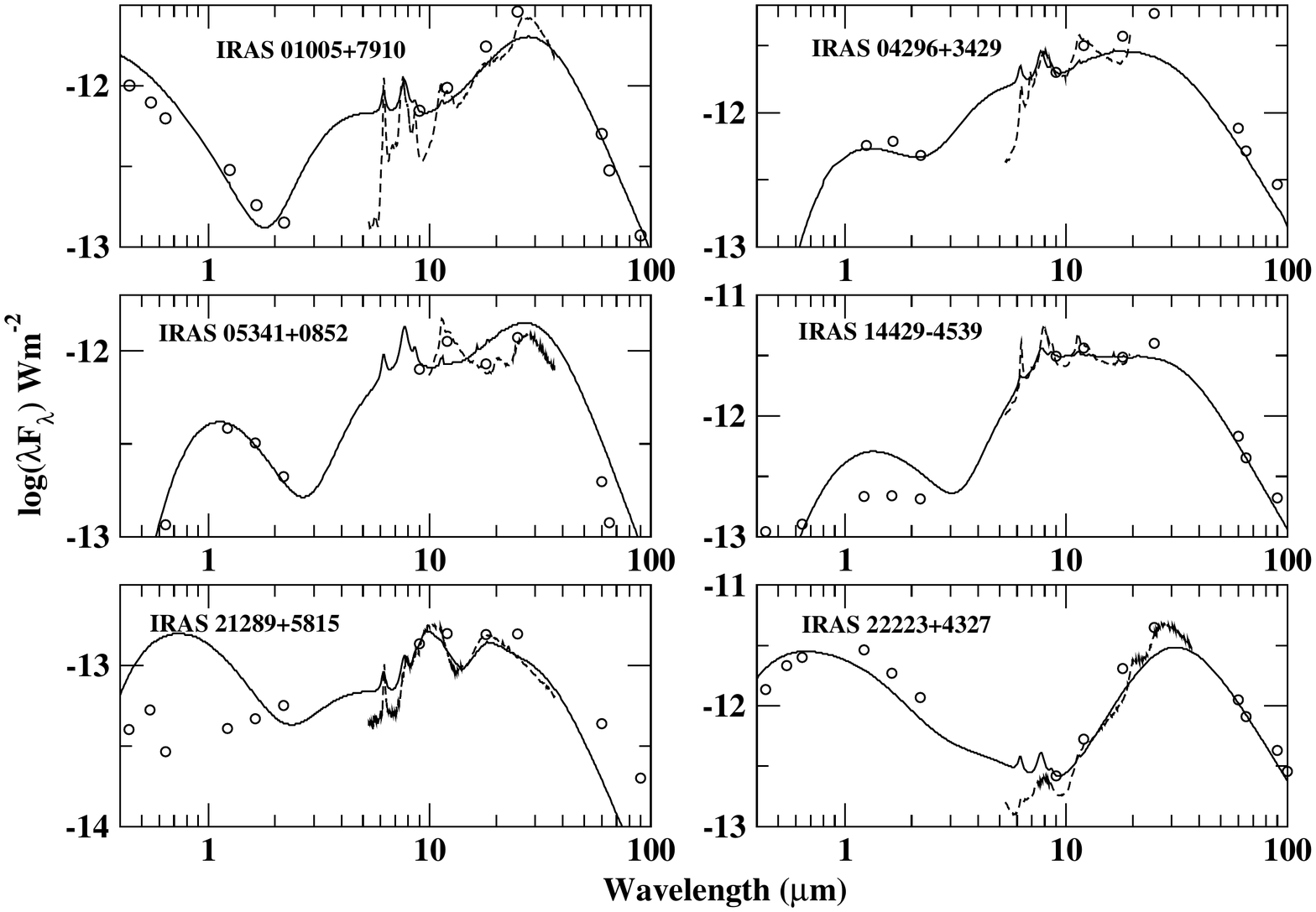}
\end{center}
\caption[Typical DUSTY models with double spherical shells for sources with PAH emission]{Model SEDs with two spherical shells using DUSTY (solid line) for a few PAGB sources that showed strong PAH emissions compared with observed data from literature (open circles). The ${Spitzer}$ IRS spectra (dashed line) are shown for comparison. Being
   in the log scale the error bars are of the point size.}
\label{ppne_dusty_v3}%
\end{figure} 

In general, the double-humped SEDs of PAGB stars indicate
detached dust shells, in contrast with the AGB stars (\citet{Venkataraman08}).
For the purpose of modeling, we have divided our sample stars into two
categories, based on their IRS spectra: (i) those that show
clearly the PAH features in the 6 - 15 $\mu$m region (ii) those showing no PAH feature at all.
Strong \& weak/blended PAHs were modeled using a double shell geometry 
because PAH alone could not account for the continuum emission on either side of
the 5 - 15 $\mu$m region, while the sources
with silicate in emission/absorption, and broad 11 $\mu$m feature were modeled using single shell geometry. 
A SED model using a single dust shell with graphite, amorphous carbon, SiC
and silicate dust grains or a combination of these, as the case may be, better represented the
circumstellar sources without PAH emissions. SED models using two shells better modeled strong, weak/blended PAH and 
mixed chemistry sources, with an inner shell of PAH and an outer shell containing 
graphite or amorphous carbon or silicate dust grains.
The optical constants for the PAH were provided externally into DUSTY while the other dust grains used here 
are in-built with DUSTY. 
A modified MRN (\citet{Mathis77}) dust size distribution was assumed 
with an exponent of 3.5 with minimum and maximum dust grain sizes of 0.005 and 0.25 $\mu$m respectively.
A radiatively driven wind model was chosen for the dust shell with density distribution of 
type 3 (see \citet{Ivezic95}), 
in which the envelope expansion is 
owing to the radiation pressure on dust grain particles
(see \citet{Ivezic97}).  
Table \ref{pagb_single} shows the list of single-shelled PAGB sources 
along with model-derived photospheric and circumstellar shell parameters, namely, T$_{*}$, T$_{d}$, \.{M},  
$\tau_{0.55}$, r$_{1}$, 
r$_{1}$/r$_{c}$ and the grain composition (Gr is for graphite and AmC for amorphous carbon and Sil for silicate grains with 
the portion in 1 given in parentheses).
Also listed in the Table 3 are the dust mass loss rates \.{M$_{d}$} using the
following empirical relation by \citet{Lagadec08},
\\
\begin{equation}
log(\dot{M}_{d})=-9.58+0.26(K_{s}-[11])+0.05(K_{s}-[11])^2-0.0053(K_{s}-[11])^3
\end{equation}
\\
where $K_{s}$ and [11] are the magnitudes of the sources
in the 2MASS K band and VISIR PAH2 filter ($\lambda_c$= 11.25 $\mu$m, $\Delta\lambda$=0.59 $\mu$m) respectively.
The dust mass loss rates obtained from infrared colors were found to be in the range 
between 10$^{-8}$ to 10$^{-10}$ M$_{\odot}$yr$^{-1}$ agreeing well with the results of \citet{Lagadec08}. 
It is clear from the table that the grain composition for all the sample stars 
is carbonaceous dust (either graphite or amorphous carbon or a combination of the two).
This result, however, does not fall in line with the finding of \citet{Cerrigone09}, whose
sample showed only 25\% having carbon rich envelopes.  
Figure \ref{ppne_dusty_single} shows typical model fits
for the PAGB stars with single shell.  
Photometric observations as well as ${Spitzer}$ IRS spectra are overplotted for comparison.

For those sources (27 in number) in which the $Spitzer$ IRS spectra showed strong PAH emissions, we have used two spherical shells
(instead of one used before) for the SED modeling using DUSTY.
The first or inner shell contains predominantly neutral or 
ionised PAH and the second, outer shell contains any of
the common dust grains (amorphous carbon/silicate/graphite). 
The ratio of the 6.2 or 7.6 $\mu$m PAH over 11.2 $\mu$m feature can be used as a 
measure of the ionization fraction of the PAHs, because the
ionized PAHs emit more strongly in the 6.2 and 7.6 $\mu$m bands while the 11.2 $\mu$m band is 
 stronger in neutral PAHs.
This ratio may be estimated from $Spitzer$ mid-IR spectra and whether the PAHs are neutral or ionized
may be ascertained in a given source.
The optical properties obtained from \citet{Li01} for neutral and ionised PAHs with grain size of 0.01 $\mu$m was used
as external input to the DUSTY software. 
The density distribution falling off as r$^{-2}$ was used for both the dust shells.
Fig \ref{ppne_dusty_v3} shows typical model fits with double shells for the PAGB candidates.  
Table \ref{ppne_2shells} show the list of these PAGB sources
along with the derived photospheric and circumstellar shell parameters, T$_{*}$, T$_{d}$, ionised state of PAH in the first sphere,
r$_{1}$, r$_{1}$/r$_{c}$ for both the shells
and the grain composition in the dust shell 2.
It is assumed that PAH and dust exist in two separate shells with the former closer to the star than the latter.
The PAH emissions may be from regions very close to the edge of the photo-dissociation regions.

We note here that it may be quite reasonable to assume that PAH molecules exist 
in a separate shell closer to the star because they require soft-UV radiation for excitation.
Further, it is possible that the PAH shell could form close to the central star
during the most recent PAGB mass loss episode where the outer layer of the central star is enriched 
with carbon rich materials due to a thermal pulse. As mentioned earlier, \citet{Waters98} used similar arguments
for explaining the observed the oxygen and carbon rich signatures in the binary system of Red Rectangle. 
As mentioned earlier, \citet{Guzman15} suggested that in the planetray nebula 
BD +30$^{\circ}$ 3639, the carbonaceous dust (PAH) exists in the inner regions while the oxygen-rich dust 
(silicates) in the outer regions of the nebula. 

\section{Results and Discussion:}

\subsection{Dust formation distance}
\begin{figure}
   \centering
 \includegraphics[width = 5.0in]{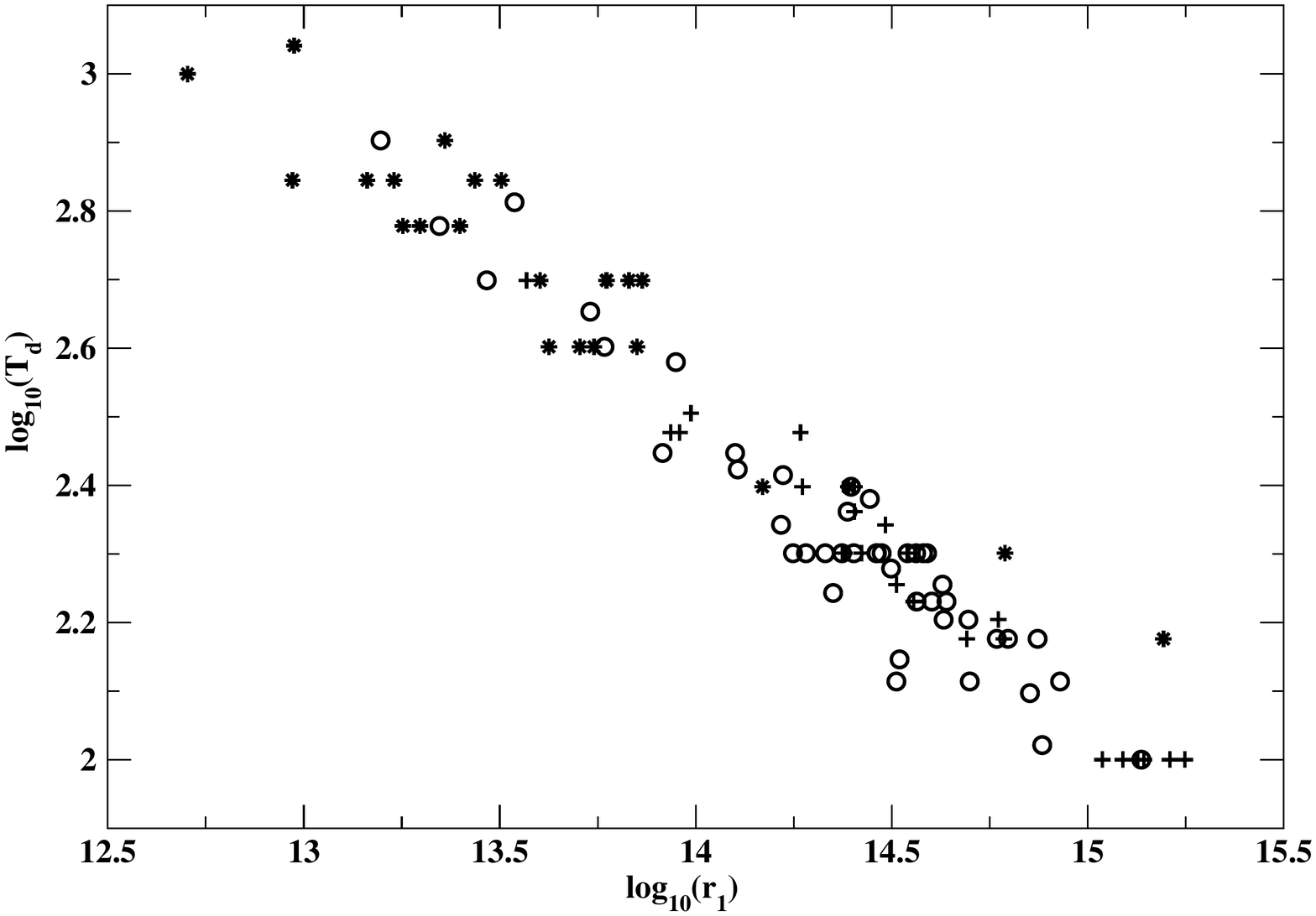}
   \caption{Plot of dust temperature T$_{d}$ 
      versus the inner dust shell radius r$_{1}$ (in m) for 
         the PAGB stars (open circles) obtained from DUSTY SED models with single shell irradiated by radiatively
         driven winds. The asterisk and pluses represent the sphere 1 (PAH) and sphere 2 for PAGB candidates modeled with
         a double circumstellar shell geometry.}
  \label{fig4}
   \end{figure}

The SED modeling reveals a few interesting, physically viable relationships between various derived parameters. 
Fig \ref{fig4} shows the plot between inner dust temperature and the inner radius of the dust shell for all the stars in our sample.
In the case of objects showing PAH spectra, the inner radii of both the PAH and dust shells are shown in the figure.  
The plot shows that as the dust temperature increases the radius of the inner shell decreases. This simply implies that 
the dust gets cooler as it is located farther from the star.
It may be inferred from the figure that 
the inner radius of the dust shell corresponding to a typical condensation temperature 
for (silicate) dust grains, T${_c}$ $\sim$ 1600K, turns out to be $\sim$ 3.5 $\times 10^{12}$ m.  
It indicates that the dust formation distance
from the stellar photosphere is typically a few stellar radii. It was argued by 
\citet{Woitke96} \& \citet{Sedlmayr00} that for
pulsationally driven mass loss to be most efficient, the dust formation should occur
at a few stellar radii (R$_{*}$).

\begin{figure}
   \centering
 \includegraphics[width = 5.0in]{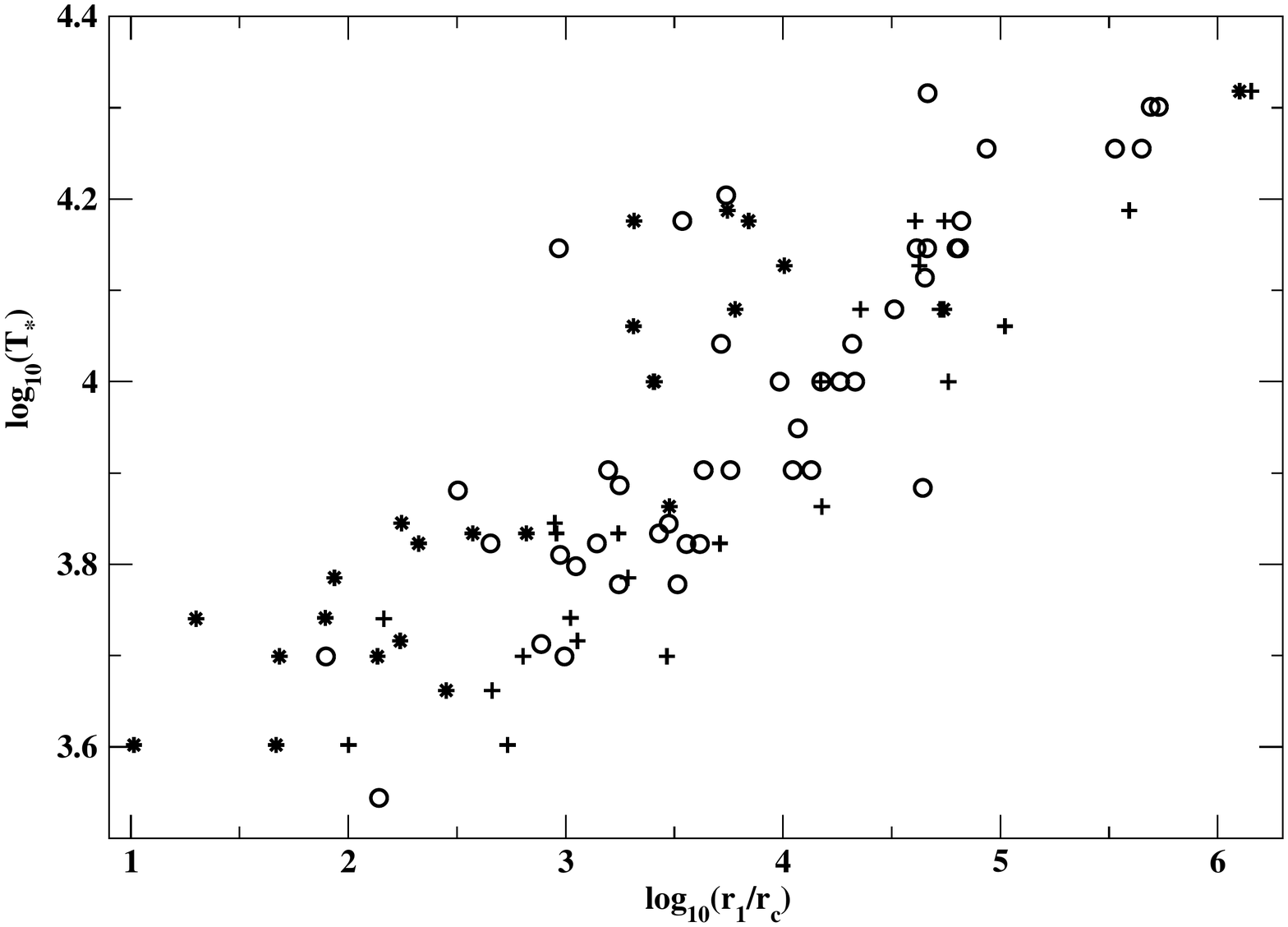}
   \caption{Plot of central source temperature T$_{*}$ versus the ratio of inner dust shell radius r$_{1}$ 
         and the central source radius r$_{c}$ for PAGB stars (open circles) obtained from DUSTY SED models with single shell irradiated by radiatively driven winds. The asterisks and pluses represent the sphere 1 (PAH) and sphere 2 for PAGB candidates modeled with a double shell geometry.}
              \label{fig5}
   \end{figure}

Fig \ref{fig5} shows the plot of T$_{*}$ as a function of ratio r$_{1}$/r$_{c}$. 
The plot indicates that the ratio increases with T$_{*}$. 
As the photospheric temperature increases the star emits copious amount of
UV photons that destroy the dust and hence the dust can survive only at distances far off from the star. This explains the   
positive correlation between the two parameters. This is in line with the above argument on
minimum distance of dust formation.

\subsection{Trends in the mass-loss rates}

\begin{figure}
   \centering
 \includegraphics[width = 5.0in]{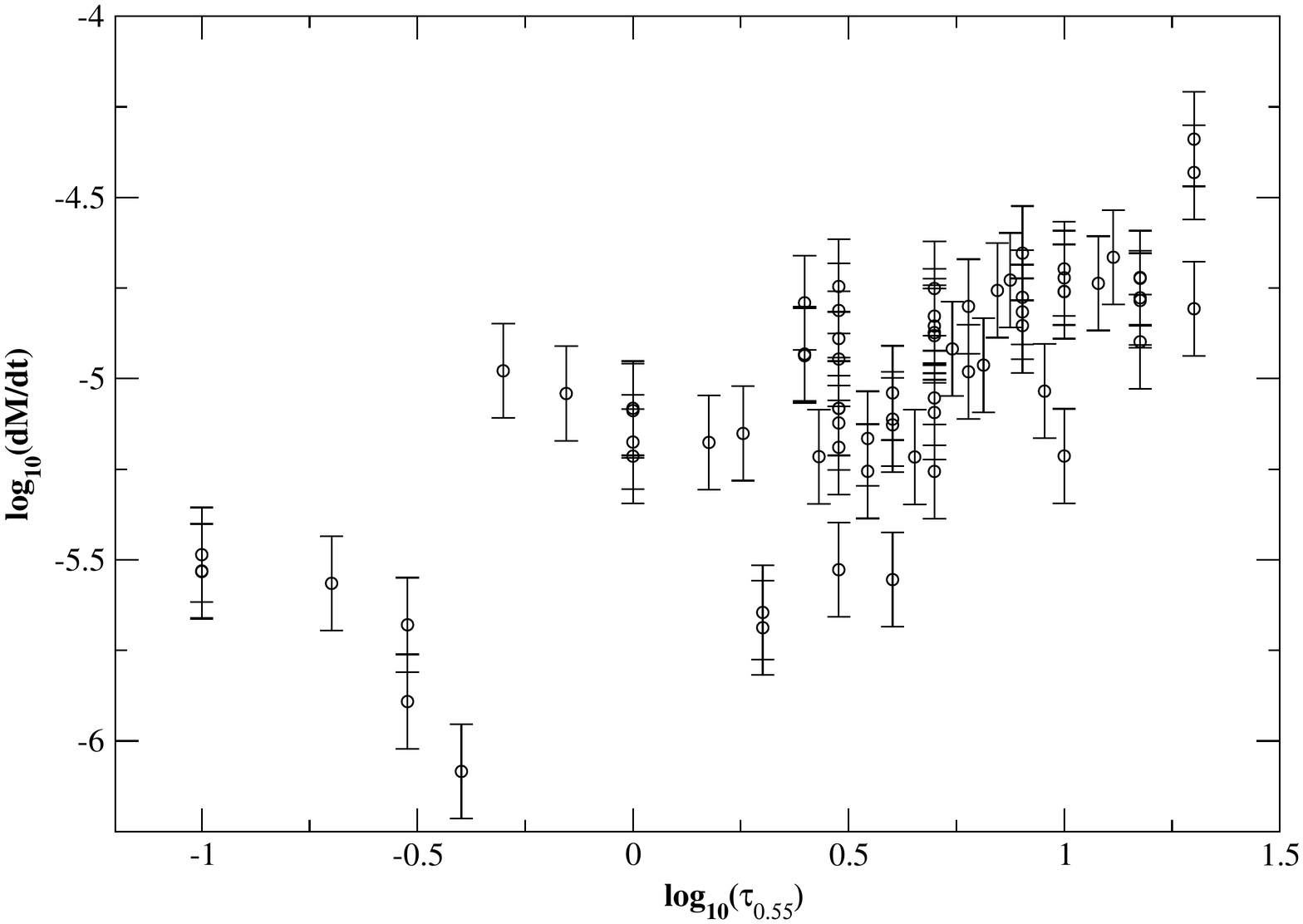}
   \caption{Plot of total mass loss rate dM/dt versus the optical depth at 
   0.55$\mu$m $\tau_{0.55}$ for PAGB stars.}
              \label{fig7}
   \end{figure}

\begin{figure}
   \centering
 \includegraphics[width = 5.0in]{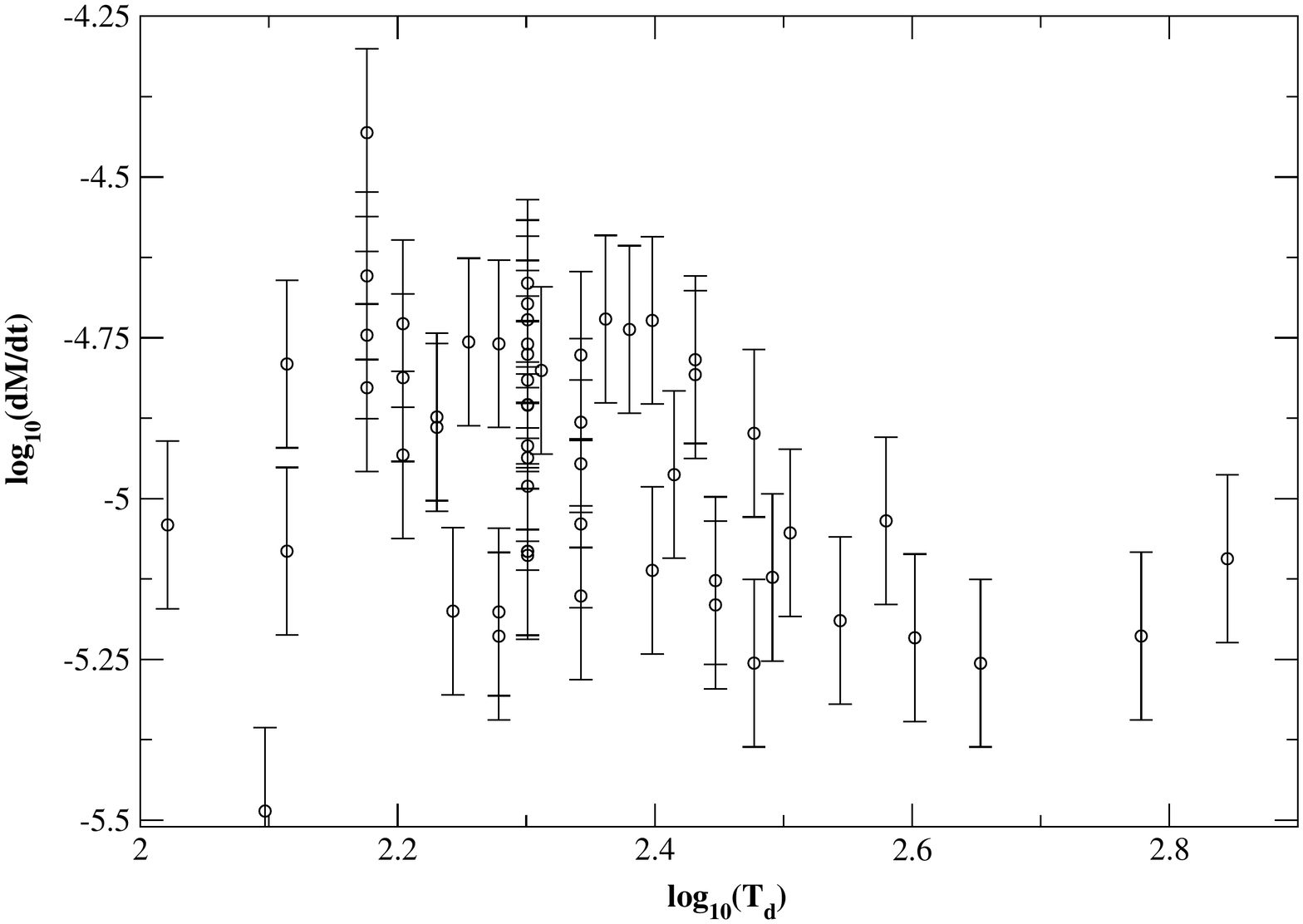}
   \caption{Plot of total mass loss rate dM/dt versus the dust temperature T$_{d}$ for 
   PAGB stars.}
              \label{fig6}
   \end{figure}

The PAGB mass loss mechanisms are not well-understood and still being debated.
Initially a continued or extended superwind phase until the central star attains a certain temperature was
assumed to explain the PAGB mass loss rates (\citet{Schoenberner83}).
However, very recently \citet{Hrivnak15} attributed the variation in the PAGB light curve
to pulsations that levitate the circumstellar envelopes to sufficient heights
leading to dust formation. As the central stars evolve and the temperature increases,
radiation driven winds facilitate the mass loss process (\citet{Vassiliadis94}). 
The mass loss rates obtained from the model for our sample of PAGB stars showed
expected trends with the other model-derived parameters.
Despite considerable scatter, especially for optically thin cases ($\tau_{0.55}$ $\le$ 1.0),
one may infer from Fig \ref{fig7} an increasing trend in mass loss rate with optical depth at 0.55 $\mu$m.
During the mass loss process, the dust shell remains initially optically thin,
allowing the radiation from the central star to escape out. But as
the mass loss increases, the optical depth increases leading to the formation of 
thick circumstellar shell resulting in formation of dust that completely blocks the radiation escaping 
the central star. 
These are physically 
expected results and hence the plots serve as a validation of circumstellar dust shell properties.
Fig \ref{fig6} shows the tendency of decreasing mass loss rate with increasing inner dust temperature
in line with the above arguments.
Since most of the dust grains (silicates and graphites) condense
 at approximately at the same temperature around 1500 K (\citet{Salpeter77}), a strong dependence on the effective temperature ($T_{*}$)
is expected. An increase in the stellar temperature moves the dust nucleation region outward and
hence towards lower densities which leads to an ineffective coupling between the gas and dust thereby
decreasing the mass-loss rates (\citet{Lamers99}). Also, as 
T$_{d}$ increases the dust shells tend to be located closer to the central star where the dust is destroyed,
leading to a low mass loss rate.

\begin{figure}
   \centering
   \includegraphics[width = 6.0in]{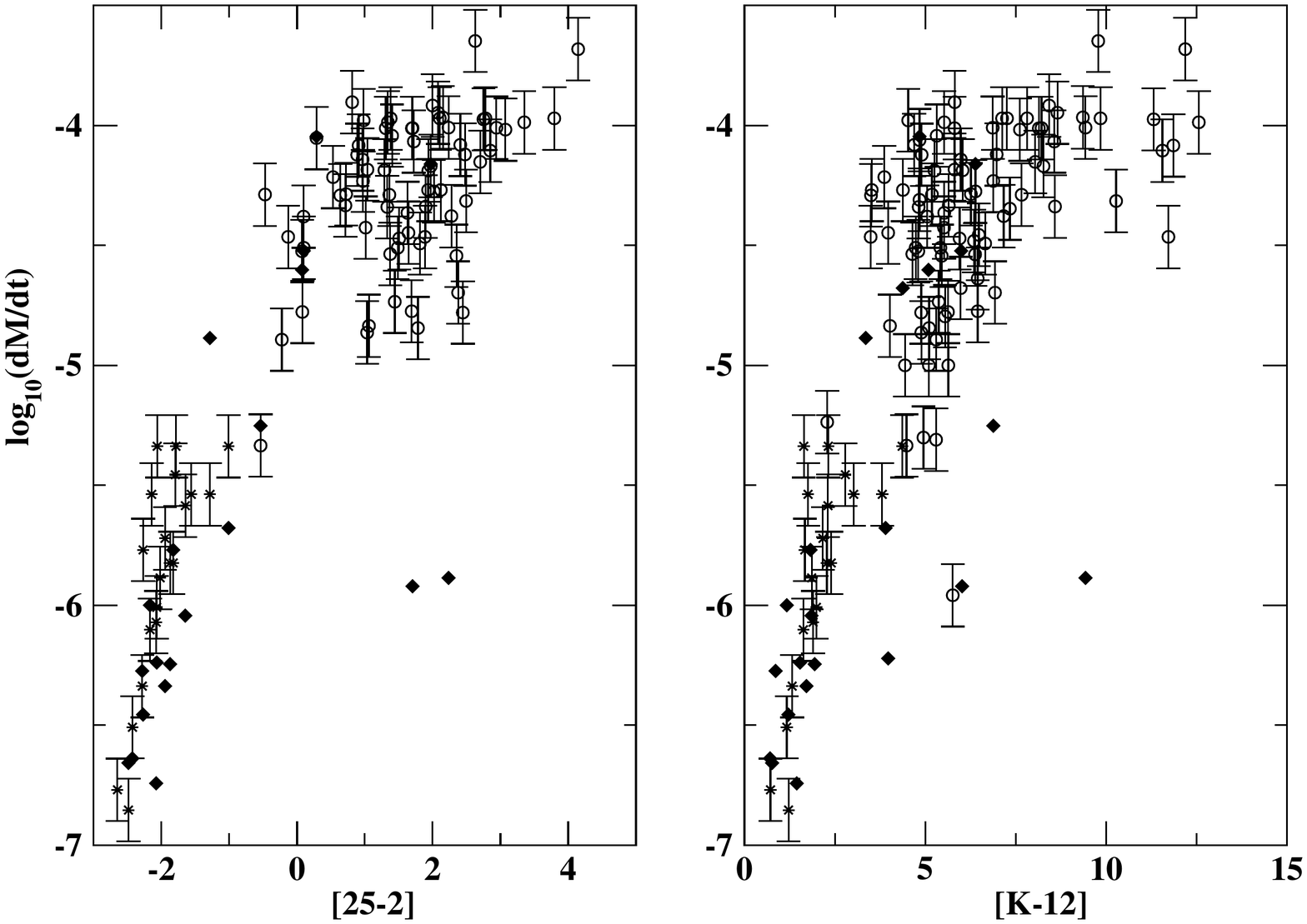}
      \caption[Plot of total mass loss rate vs mass loss index]{Plot of the total mass loss rate obtained from DUSTY
       with the mass loss index [25-2] \& [K-12] for AGB (asterisks) and PAGB stars (open circles). The mass loss rates
       derived from the observed rotational transitions of CO at mm wavelengths for some of the AGB and PAGB stars in our sample (filled diamonds) are also shown for comparison.}
   \label{mass_loss_mtype_pagb}
   \end{figure}

\begin{figure}
   \centering
   \includegraphics[width = 6.0in]{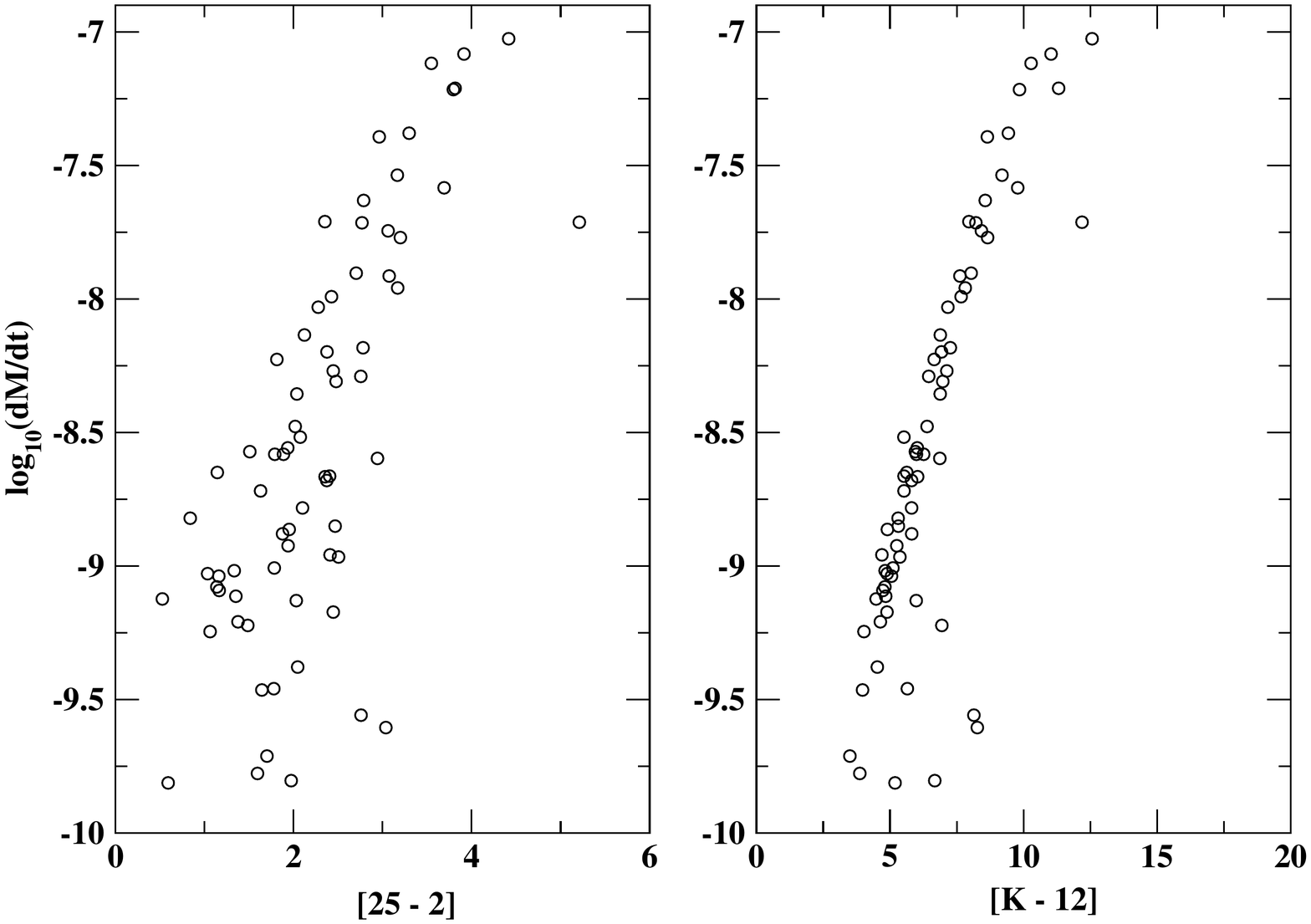}
      \caption[Plot of  dust mass loss rate vs mass loss index]{Plot of the dust mass loss rate 
      obtained using the relation from \citet{Lagadec08}
       with the mass loss index [25-2] \& [K-12] for the sample PAGB stars (open circles).}
   \label{mass_loss_mtype_pagb1}
   \end{figure}

The color [K-12] is often considered as an indicator of mass loss in AGB/PAGB stars (\citet{Lebertre98}). 
\citet{Anand93} proposed that the color index [25-2] as a better tool for representing mass loss rate in AGB stars.
Fig. \ref{mass_loss_mtype_pagb} shows the plot of the total mass loss rates 
for all those PAGB stars that have single dust shells with [25-2] \& [K-12].
Also included in the figure are the sample of AGB stars from our previous work (\citet{Venkataraman08}). 
The figure indicates that the mass loss rate does not cease completely towards the end of the AGB phase
but extends beyond and towards the PAGB phase. 
The PAGB mass loss rate was found to be much higher than that during the AGB phase.
Figure \ref{mass_loss_mtype_pagb} also shows the mass loss rates for some PAGB stars (common with our sample) 
determined from the observed CO rotational line profiles by several authors, namely 
\citet{Woodsworth90}, \citet{Likkel91},
\citet{Omont93}, \citet{Hrivnak99}, \citet{Bujarrabal01},
\citet{Hoogzaad02} and \citet{Hrivnak05}.
One can see a good agreement between the model values and those determined by observations.
Fig \ref{mass_loss_mtype_pagb1} shows the plot of dust mass loss rates as a function of the colors 
[25-2] and [K-12], indicating a similar trend as the total mass loss rate. 
Fig \ref{mass_loss_mtype_pagb} shows
two distinct regions: an initial positive correlation of mass-loss rate with [K-12] \& [25-2] colors 
due to the initiation of the superwind towards the end of AGB phase, and 
the continued constant high mass loss rates during the PAGB stage possibly 
due to the pulsation and radiatively driven winds from the central star.
Recently, \citet{deVries14} suggested an intense mass loss due to 'hyperwinds'
that occur between the superwind and the PPNe phase to explain
the high mass loss rates during the PAGB phase. 
It is, however, not clear as to what causes this hyperwind, or for that matter, the superwind itself.
\citet{deVries15} also
suggested the presence of crystalline forsterite (Mg$_{2}$SiO$_{4}$) as a 
signature for massive AGB stars. 
Based on the variation in the circumstellar nebular conditions and the 
observed variations in the spectral features, \citet{Garcia03} had proposed an evolutionary scheme 
for both oxygen and carbon rich sources in transition from AGB to PN phase.
About 15 stars in our sample show the presence
of crystalline forsterite at 33.6 $\mu$m feature suggesting possibly their massive nature.
These massive AGB stars cannot lose their entire envelope during
the superwind phase alone. So the intense mass loss should continue 
beyond the superwind phase before becoming a PN, thus explaining
our SED model results.
\citet{Hrivnak05} had suggested an intense mass loss during the PAGB phase
responsible for the steeper density law for fitting 
their observed CO rotational line profiles. 
  

 \subsection{Trends in PAH modes with the evolution:}
  
   \begin{figure}
    \centering
 \includegraphics[width = 5.0in]{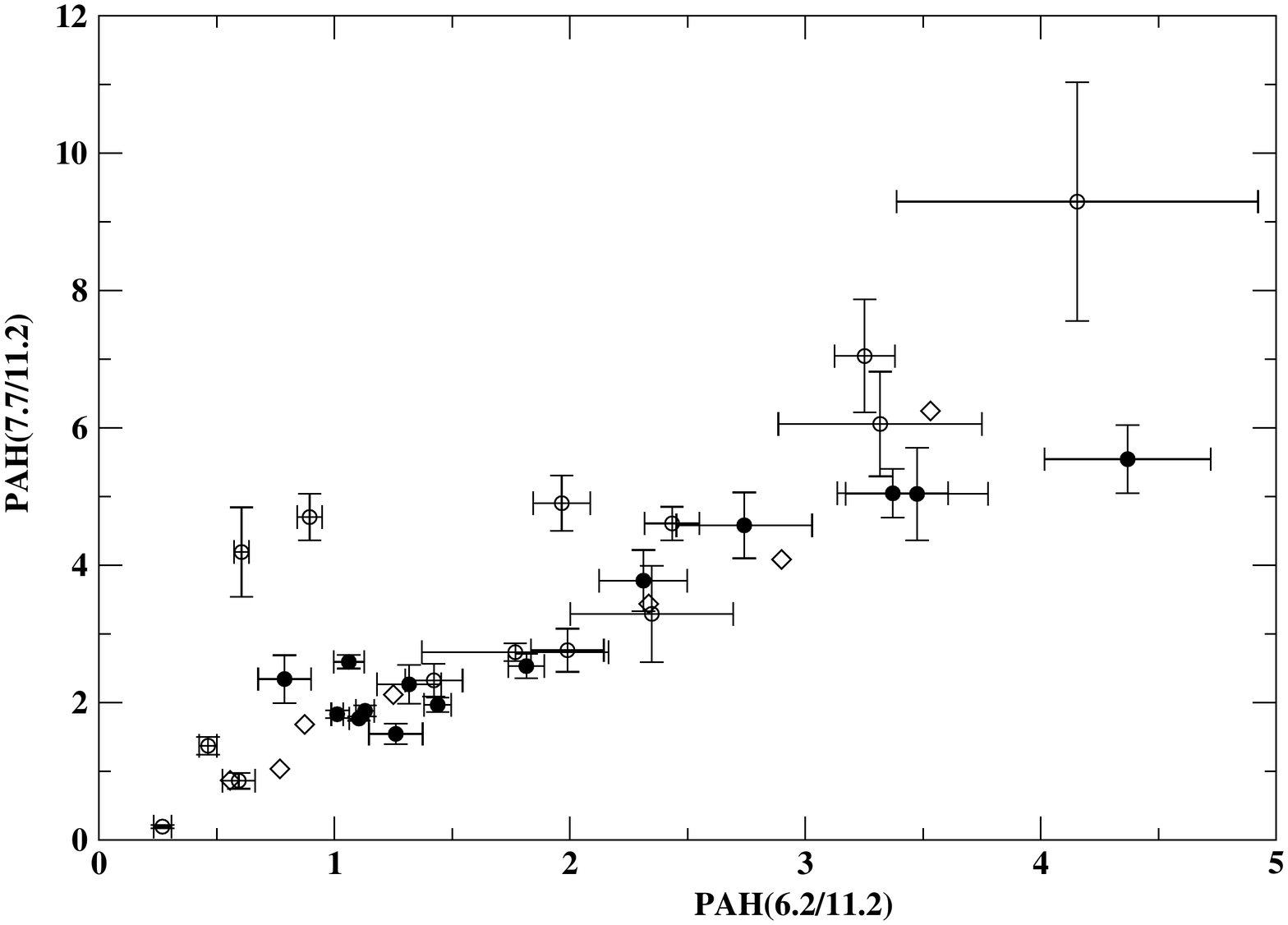}
    \caption{Plot of PAH mode flux ratio [7.7]/[11.2] vs [6.2]/[11.2]  for 
    PAGB stars. The open circles represent the PAGB stars in our sample while the filled circles represent
    PNe in the Large Magellanic cloud (LMC) and Small Magellanic cloud (SMC) and the open diamonds represent the same 
    in our milky way galaxy obtained from \citet{Bernard09} are shown for comparison. }
               \label{fig10}%
     \end{figure}
 
   \begin{figure}
   \centering
 \includegraphics[width = 5.0in]{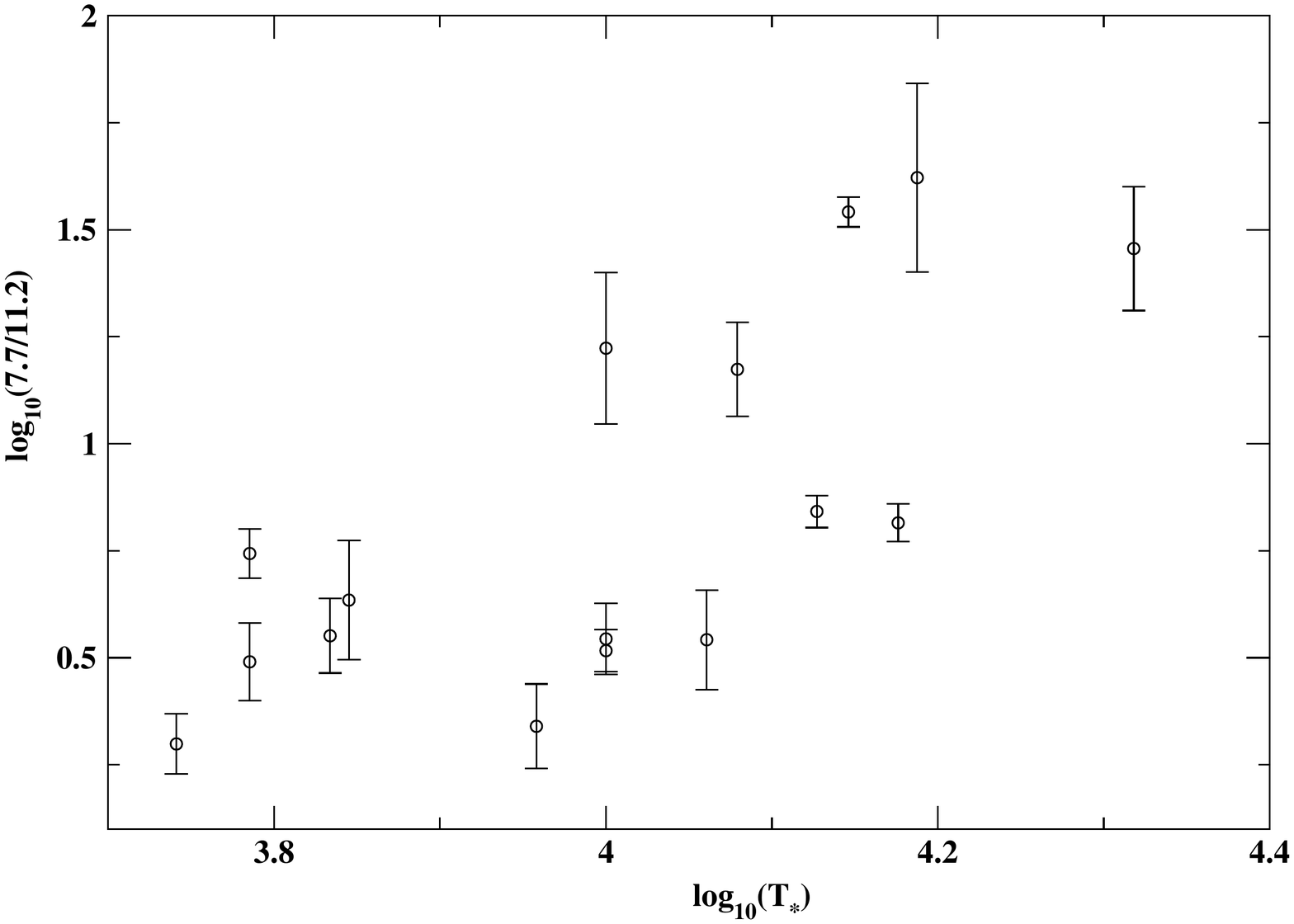}
   \caption{Plot of PAH mode ratio EW[7.7]/EW[11.2] versus 
   the effective temperature T$_{*}$ for PAGB stars.}
              \label{fig8_1}%
   \end{figure}

\begin{figure}
   \centering
 \includegraphics[width = 5.0in]{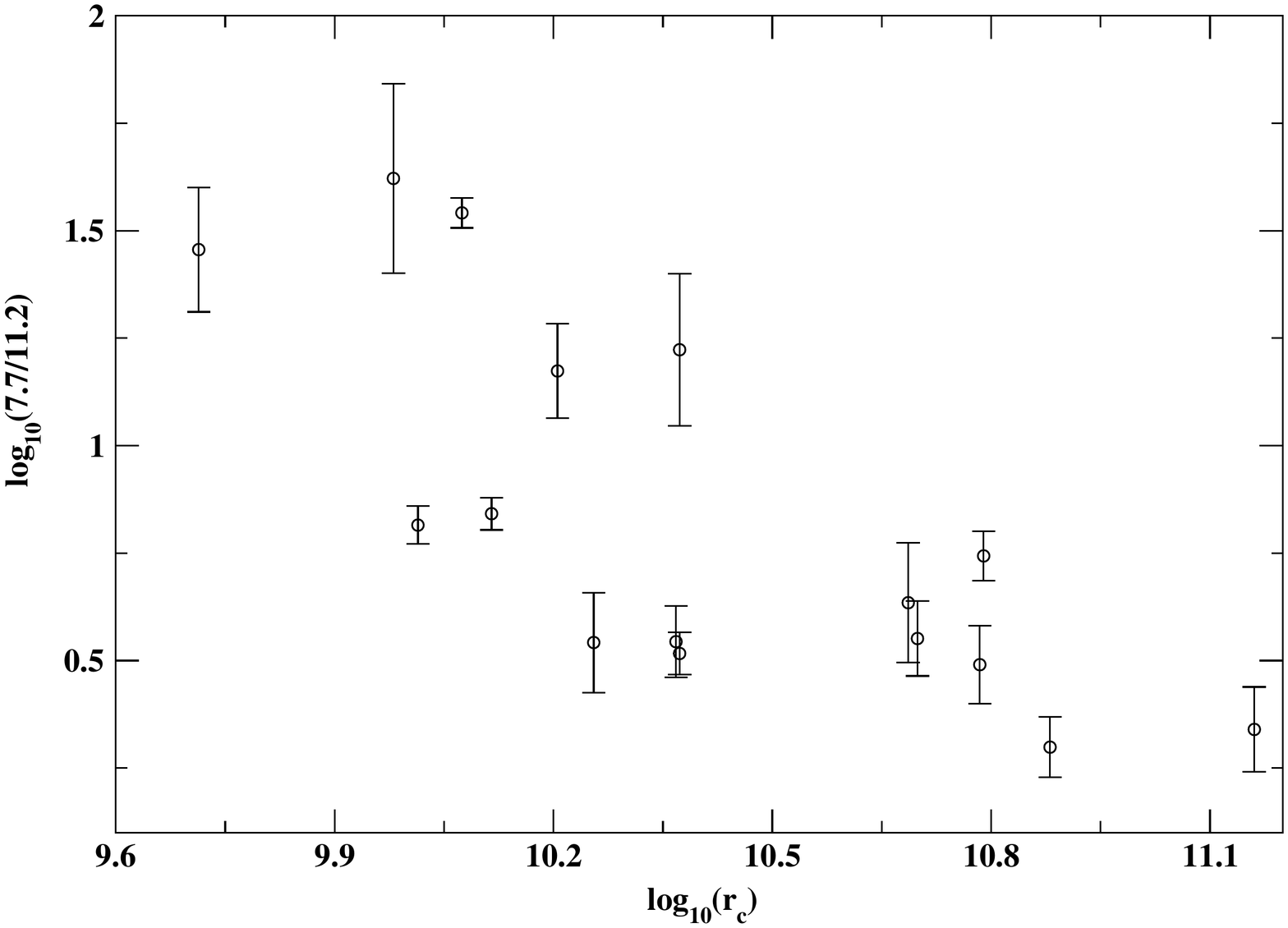}
   \caption{Plot of PAH mode ratio EW[7.7]/EW[11.2] versus 
   the radius of the central source r$_{c}$ for PAGB stars.}
              \label{fig8_2}%
   \end{figure}

   \begin{figure}
   \centering
 \includegraphics[height = 5.0in, width=5.0in]{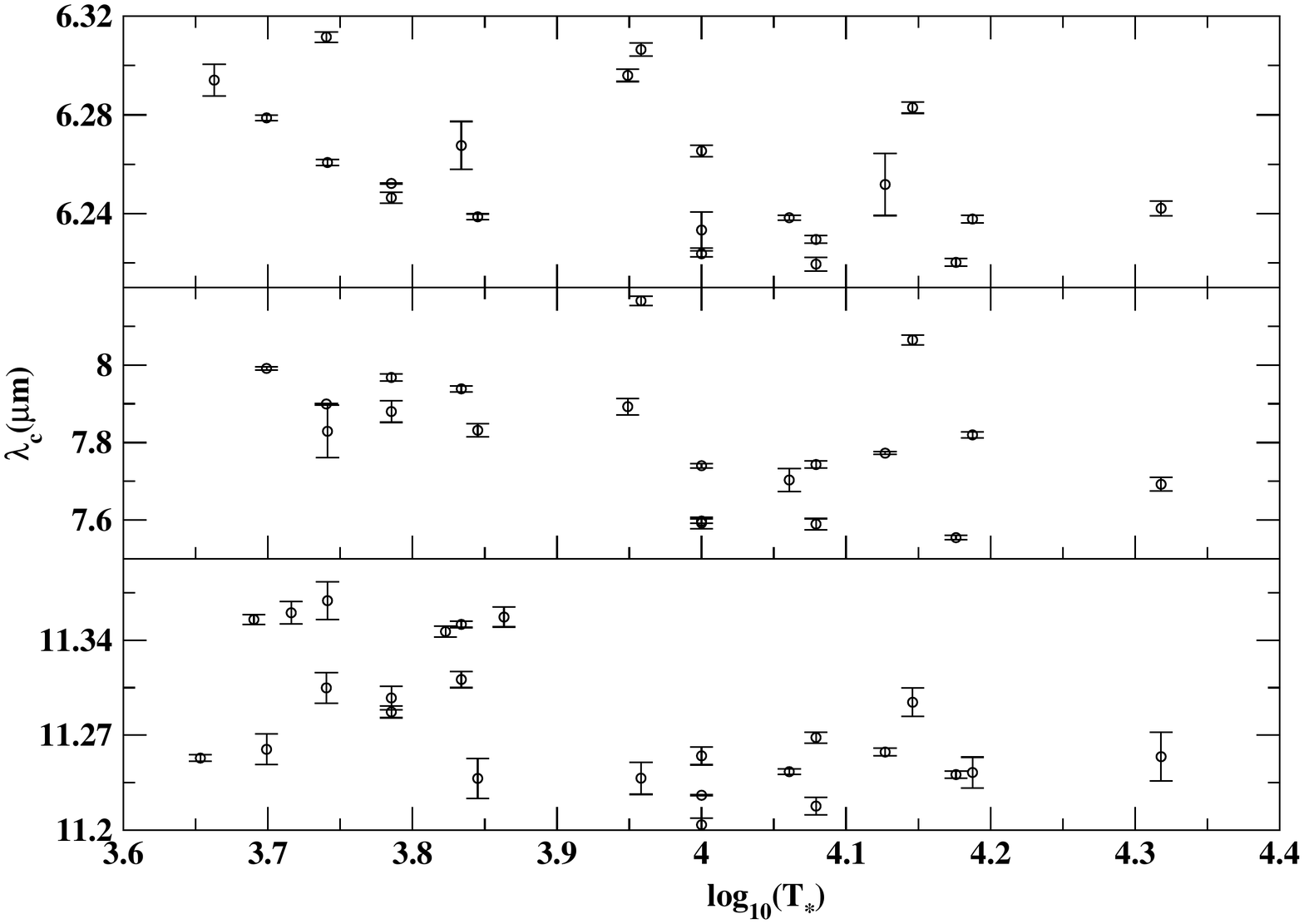}
   \caption{Plot of the peak wavelength of 
   the PAH feature at 6.2, 7.7 \& 11.2 $\mu$m versus the photospheric temperature T$_{*}$ 
   for PAGB stars.}
              \label{fig11}%
   \end{figure}

Here we examine the trends in the strength of the PAH emission modes with the stellar effective temperature T$_{*}$.
The T$_{*}$ is used to represent the evolutionary stage of the sources considered 
in the present study; higher the value, the more advanced is the evolution of the source. 
Interesting correlations were obtained between the strengths of the PAH spectral features 
 with the physical and circumstellar parameters of the PAGB candidates by \citet{Peeters02},
 \citet{Sloan05}, \citet{Bernard09} and \citet{Hony01}.
In this section, we examine the present sample of PAGB objects for 
such possible correlations with stellar temperature and other 
parameters that characterize circumstellar envelope.

Firstly, we examined the correlation between intensity ratio I(7.7)/I(11.2) and I(6.2)/I(11.2). 
Each of these ratios is known to indicate the ionization fraction in PAH molecules (\citet{Galliano08}). 
Fig \ref{fig10} shows a plot of these intensity ratios indicating a positive correlation, 
a trend that was earlier shown by \citet{Cerrigone09}.

We have also studied the relative dominance of C-C stretching and C-H in-plane bending modes (e.g. 7.7 $\mu$m)
and C-H out-of-plane (e.g. 11.2 $\mu$m) modes as the stars evolve.
It may be noted that we have used the equivalent width (EW) as the measure of the strength
of the PAH feature instead of the feature intensity used by other authors 
(e.g. \citet{Cerrigone09}, \citet{Zhang10}).
We have found that using EW yielded essentially the same results as using
intensity of spectral features.  
Fig \ref{fig8_1} and \ref{fig8_2} show the ratio of the EWs of the PAH features  
 [7.7]/[11.2] versus T$_{*}$ and radius of the central source r$_{c}$ respectively. 
The PAH feature at 7.7 $\mu$m represents the C-C stretching and C-H in-plane bending modes 
  and the 11.2 $\mu$m feature represents the solo C-H out-of-plane bending modes. 
  One may notice a tendency of increasing ratio with T$_{*}$ 
 indicating the dominance of C-C stretching and C-H in-plane bending modes 
 as the objects evolve; while a decreasing tendency may be noticed with r$_{c}$.
As mentioned earlier, one may also conclude from this plot that stars showing
 ratio [7.7/11.2] $>$ 1 have higher fraction of ionised PAHs than those with lower ratio.
 
In their study of a few PAGB candidates, \citet{Zhang10} showed that the ratio  
 [12.6/11.2] tends to decrease with  
 the central star temperature, indicating that the solo C-H modes 
 at 11.2 $\mu$m become more and more predominant over the duo- or trio- or quarto- modes at 12-13 $\mu$m. This 
 is suggestive of the possibility that the solo C-H out-of-plane bending modes 
 get stronger as the star evolves to PNe (\citet{Kwok04}). 
 We also found in the present sample that the ratio indicated a decreasing trend with T$_{*}$ (not shown here).
 The ratio is less than 1 for nearly all the sources that 
 showed PAH bands suggesting that the solo-modes dominate over the duo and trio modes. 
 Our result is in agreement with \citet{Zhang10}.
 In most cases of PNe in their sample, Hony et al (2001) found that the ratio as a function of hard radiation 
 field remained less than 1.    
 
We have also seen if there is a trend between the central source temperature and  
the plateau region 16-20 $\mu$m representing the blended features of in-plane or out-of-plane
C-C-C bending modes normalised with the 11.2 $\mu$m solo band. 
A decreasing trend of the mode with T$_{*}$ was noticed (not shown here).

\citet{Sloan07} showed that the peak wavelength of the 11.2 $\mu$m feature changes
 with T$_{*}$ for a small sample of carbon stars. We verified this result with our sample. 
 Fig \ref{fig11} shows the plot of the peak wavelengths of the 6.2, 7.7 and 11.2 $\mu$m features with 
 T$_{*}$ (see Table 2). 
 One may notice, in general, a shift towards the shorter wavelengths
 with increasing T$_{*}$, up to a value of 10$^{4}$ K in agreement with \citet{Zhang10}.
\citet{Sloan07} assume that PAH clusters are embedded in a matrix of
aliphatic groups. When such hydrogenated amorphous carbon (HAC) material is exposed to hard radiations, the
aliphatic bonds break down progressively leading to the positional shifts of PAH features (for details see \citet{Sloan07}).

\section{Identification of transition objects}
\begin{figure}
\centering
\includegraphics[width = 6.0in]{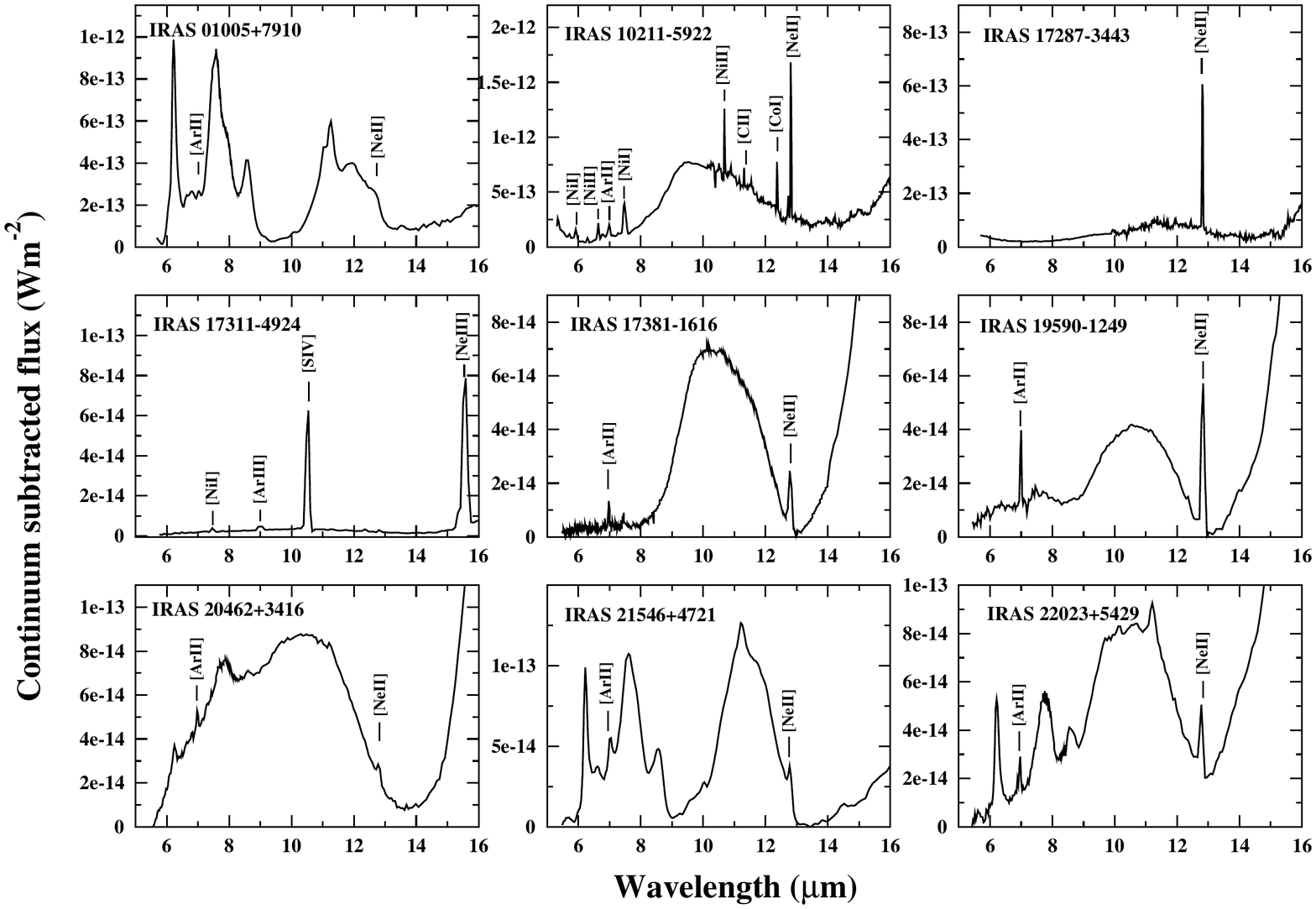}
\caption[{Transition objects with [NeII] emission at 12.8 $\mu$m}]{Transition objects from our sample 
   showing prominent low and high excitation fine structure lines}
\label{transition_ne}%
\end{figure}
The transition process from the post-AGB to PNe phase still remains elusive in stellar evolution.
After the termination of the high AGB mass loss, the circumstellar envelopes expand and cool,
and the effective temperature of the central star increases to ionize the
circumstellar material towards the end of the post-AGB phase 
(\citet{Kwok81}, \citet{Kwok84}).
Mid-IR fine structure lines suffer less dust extinction compared to the optical and UV emission lines
and hence are easier to detect.
However not all the PAGB objects show these emission lines that indicate their 
transition nature; a possible explanation is in order.
The morphologies of PAGB dust shells are believed to be predominantly axisymmetric
(\citet{Ueta00}) that interact with the PAGB winds to form aspherical PN shells. 
Mid-infrared imaging of PAGB sources show a central dust torus around the central source (\citet{Ueta01}). 
The mid-IR morphologies of the circumstellar shell are characterized into two groups namely toroidal and
 core/elliptical (\citet{Meixner99}). The toroidal dust shells have very low optical depth and the star light can scatter in all directions. 
In the case of elliptical mid-IR shells the optical depth is
high and the starlight can escape only through the bi-cone opening of the dust shell making bipolar nebulosities. 
The fine structure emission lines are believed to arise from the irradiated central torus atmospheres. In order to 
observe these lines, the inclination
of the source and the optical depth along the line of sight play a crucial role. 
If the source is viewed pole-on then the line of sight optical depth is the highest 
from the circumstellar shells and makes the detection difficult.
When observed edge-on one may view the interaction of the fast winds with the dense equatorial torus.
Hence the detection of these lines depends on the morphology and inclination of the source.

Figure \ref{transition_ne} represents $Spitzer$ IRS spectra of transition objects with hot cores ionizing the circumstellar
matter, thereby making a transition from PAGB 
to PNe phase, showing low and high excitation fine structure lines. 
There are around 31 objects in the sample which may be termed as hot objects having
T$_{*}$ $\geq$ 10$^{4}$ K but only 9 of these show fine structure lines of Ne.
The IRS spectra of the PAGB candidate IRAS 01005+7910
IRAS 21546+4721 and IRAS 22023+5429 showed PAH emission in addition to [NeII].  
Among the other PAGB candidates showing transition behavior,
IRAS 10211-5922, IRAS 17381-1616 and IRAS 19590-1249 showed silicate in emission.
The source IRAS 10211-5922 shows [NiI], [NiII], [ArII], [CoI] and [CII] in addition to the [NeII] line, thereby suggesting
its advanced stage of evolution among the other candidates.
Also, IRAS 17287-3443 and IRAS 17311-4924 show peculiar dust properties in addition to the fine structure lines. IRAS 17311-4924
shows high excitation lines of [ArIII], [SIV] and [NeIII] in addition to the [NiI] lines.
It also shows faint emission of [NeII] line at relatively lower S/N. 
Further, it shows a rising mid-infrared continuum (near the [SIV] line) possibly due to VSGs that can survive 
close to the hot source (\citet{Lebouteiller07}).
In the light of these fine structure line emissions and the contribution due to VSGs, IRAS 17311-4924
might be a high excitation transition object.  
The sources IRAS 20462+3416 and IRAS 22023+5429 are mixed chemistry objects with [NeII] emission.  
As argued above, the reason for fewer candidates showing [NeII] emission could be due to orientation effect
and deeply embedded nature of the elliptical dust shells. 
Our SED model of sources with hot central temperatures also shows large $\tau$$_{0.55}$ values.
Even with the uncertainty associated with model $\tau$$_{0.55}$ values, the trend
is compelling: those that showed [NeII] line have much lower $\tau$$_{0.55}$
values with an exception of IRAS 17287-3443 compared to those that did not.

\section{Detection of Fullerene in IRAS 21546+4721}

\begin{figure}
\centering
\includegraphics[width = 7.0in]{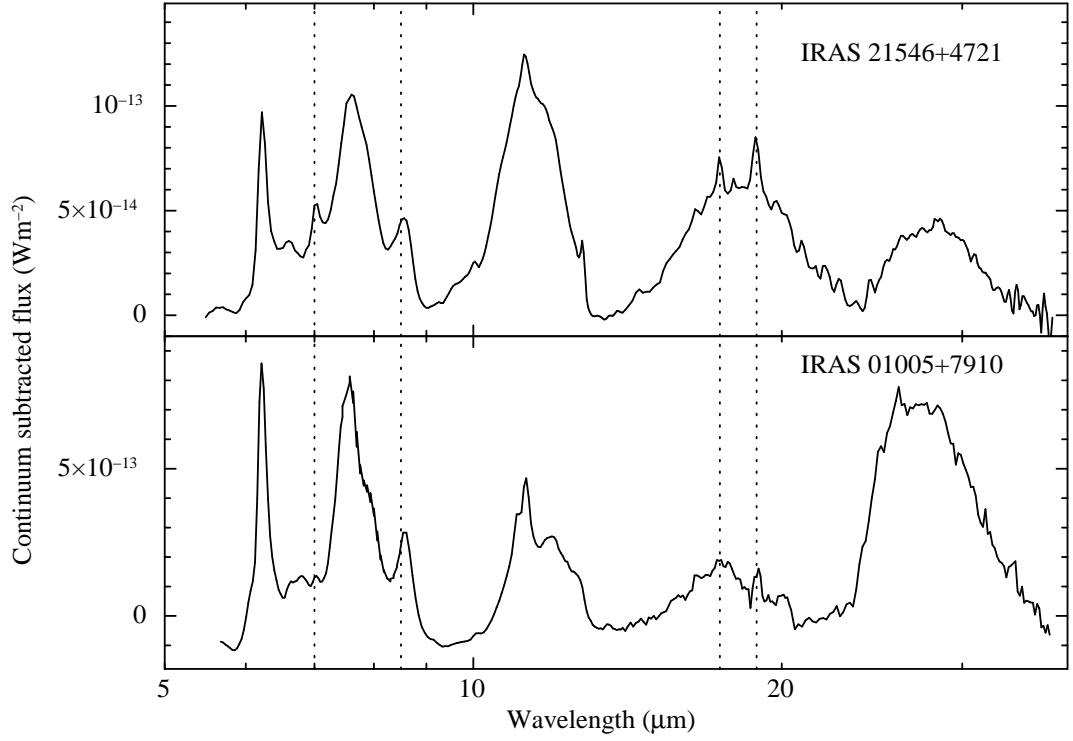}
\vspace{-20mm}
   \caption[${Spitzer}$ IRS spectra of sources with fullerene]{${Spitzer}$ IRS spectra of PAGB 
   candidates IRAS 21546+4721 and IRAS 01005+7910
   showing vibrational modes of C$_{60}$ fullerene }
\label{fullerene}%
\end{figure}
Fullerene molecules (C$_{60}$) are large molecules made of carbon hexagons 
and pentagons that are organised in the shape of a hollow
sphere (\citet{Kroto85}, \citet{Kroto88}, \citet{Curl88}, 
\citet{Kratschmer90}). 
Fullerenes are formed not only in hot and dense envelopes of evolved stars, but also
identified in the tenuous and cold environment of interstellar clouds illuminated by strong ultraviolet radiation. 
The formation mechanism of this molecule in space is still unclear.
The observed abundance of fullerene challenges the standard ion-molecule or grain-surface chemistry formation routes.
Laboratory results showed an inhibition in the fullerene formation with the presence of hydrogen,
while supporting the formation of PAH (\citet{deVries93}). 
Another mechanism that is getting much interest in recent times is the top-down model whereby a photochemical processing
of Hydrogenated amorphous carbon (HAC) grains results in the formation of PAH and fullerene (\citet{Garcia10}, \citet{Micelotta12}, \citet{Bernard12}).
The mid-IR vibrational modes of C$_{60}$ lie at 18.9, 17.4, 8.5, 7.0 $\mu$m.
The first astronomical detection of fullerene in the circumstellar environment of young PN, 
Tc1 was made by \citet{Cami10}.
Fullerene was detected in young carbon-rich PNe in Milky Way galaxy 
(\citet{Bernard12} \& \citet{Otsuka14}) and Magellanic clouds (\citet{Bernard12}).
\citet{Bernard12} suggested that such physical conditions as found in these PNe could favour for the formation of fullerene.  
C$_{60}$ had been detected in a few post-AGB binary stars earlier by \citet{Roberts12}
and \citet{Gielen11} in a mixed oxygen-carbon rich environment.
\citet{Zhang11} had detected C$_{60}$ in the PPN IRAS 01005+7910.
All in all there are very few detections of C$_{60}$ in late type stars.
We carefully searched for C$_{60}$ vibrational modes in the $Spitzer$ IRS spectra of 
PAGB candidates in our sample. We detected vibrational modes of C$_{60}$ 
for the first time in one of the PAGB candidate IRAS 21546+4721. 
To the best of our
knowledge this is the first detection of C$_{60}$ in this source and one of the very few detections reported yet. 
Figure \ref{fullerene} shows the vibrational modes of C$_{60}$ observed in IRAS 21546+4721 compared with those of IRAS 01005+7910.
The spectra of both the objects also show the C-C stretching and C-H bending modes of vibration of PAH. 
The vibrational transitions of fullerene at 17.4 and 18.9 $\mu$m are very clear in both the cases while those at 7.0 and 8.5 $\mu$m are blended with [Ar II] and PAH respectively. 
Modeling of the spectral energy distribution indicates a photospheric temperature greater than 10$^{4}$ K for both the sources. 
The presence of forbidden line of [NeII] in IRAS 21546+4721 further indicates its advanced stage of evolution.
Also both the spectra show strong emissions of 30 $\mu$m feature (see Fig \ref{fullerene}) that is attributed 
to MgS (\citet{Goebel85}) 
or a carbonaceous compound (\citet{Zhang09}).  
\citet{Otsuka14} found the presence of 
30 $\mu$m feature in all their sample Galactic PNe that have fullerene.
Further, the absence of 21 $\mu$m feature 
in the mid-IR spectra of IRAS 21546+4721
and IRAS 01005+7910 rules out the possibility of fullerene as a carrier for this
feature as suggested by \citet{Justtanont96} \& \citet{Garcia99}.

\section{Conclusions:}  

 \begin{enumerate}
 
 \item Based on our analysis of $Spitzer$ mid-infrared spectra of a fairly large sample of PAGB stars, 
 we have broadly classified them in to those having
 (a) strong/weak or blended PAH (b) silicate emission, in some cases along with PAH features 
 (c) silicate absorption  (d) prominent broad bump around 11 $\mu$m and  
 (e) nearly featureless spectra.
 
 \item The SED modeling of the PAGB objects shows that the temperature of the inner dust shell decreases  
        with increasing inner radius r$_{1}$, while the ratio of r$_{1}$/r$_{c}$ shows an increasing 
        tendency with stellar photospheric 
        temperature; the total mass loss rate correlates 
        negatively with dust temperature and shows an increasing tendency with $\tau_{0.55}$.  
   
   \item The post-AGB mass loss rates obtained from the SED models are higher than those of the AGB and show correlation with the mass loss indices [K-12] and [25-2]. Dust mass loss rates obtained independently from the IR colors and $Spitzer$ spectra also followed a similar trend. 
      
        \item The PAGB objects that show PAH emission features
      in the range 5-20 $\mu$m indicate the onset of UV radiation from the central star and hence represent their
      advanced stage of evolution. 
      The ratio of the observed strength of 7.7 and 11.2 $\mu$m PAH feature showed an increasing trend with T$_{*}$, while decreasing with r$_{c}$.
      The peak wavelengths of the 6.2, 7.7 and 11.2 $\mu$m features seem to shift towards shorter wavelengths with increasing T$_{*}$.  
     
     \item Nine objects in our sample showed fine structure line of Ne, indicating their transition nature from PAGB to PNe with the onset of ionisation.
     
     \item Fullerene was detected for the first time in the PAGB star IRAS 21546+4721.

 \end{enumerate}

\section*{acknowledgments}
   This research was supported by the Department of Space, Government of India. 
   The research has made use of the SIMBAD database operated at CDS, Strasbourg, France;  
   the NASA/IPAC Infrared Science Archive, which is operated 
   by the Jet Propulsion Laboratory, California Institute of Technology, under contract 
   with the National Aeronautics and Space Administration, USA;  
   and the data products from the Two Micron All Sky Survey, 
   which is a joint project of the University of Massachusetts and the Infrared Processing 
   and Analysis Center/California Institute of Technology, funded by the National Aeronautics 
   and Space Administration and the National Science Foundation.
   We acknowledge the DUSTY team for making available their codes for 
   the astronomy community. We sincerely thank the anonymous referee for very useful comments which
   improved the quality of the paper.

\bibliographystyle{mnras}
\bibliography{venkat}

\clearpage
\begin{center}
\footnotesize
\begin{longtable}{lcc}
\caption{Source of spectra and dust classification of the PAGB objects}
\label{pagb1} \\
\hline
 Object name   & Source of spectra &  Dust classification \\
\hline
\endfirsthead
\multicolumn{3}{c} %
{\tablename\ \thetable{} -- \textit{continued from previous page}} \\
\hline
 Object name   & Source of spectra &  Dust classification \\
 \hline
 \endhead
 \hline \multicolumn{3}{r}{{Continued on next page}} \\ \hline
 \endfoot
 
 \hline
 \endlastfoot

01005+7910               & $Spitzer$ SL, LL, SH      &    PAH\\
04296+3429               & $Spitzer$ SL, SH, LH      &    PAH\\
05089+0459               & $Spitzer$ SH			   & Not classified\\
05113+1347               & $Spitzer$ SH			   &    PAH \\
05341+0852               & $Spitzer$ SH, LH			&    PAH \\
06034+1354               & $Spitzer$ SH, LH			 &    Sil emission\\
06530-0213               & $Spitzer$ SH, LH			 &    PAH \\
06556+1623               &  $Spitzer$ SL, LL			  &    Not classified \\
07134+1005               & $Spitzer$ SH, ISO			   &    PAH \\
08005-2356               & $Spitzer$ SH, LH	&  11 micron source \\ 
08335-4026              & $Spitzer$ SL, LL &   PAH+Sil \\
10211-5922              & $Spitzer$ SL, SH, LH &   Sil emission\\
11201-6545              & $Spitzer$ SL, SH, LH &   Sil emission \\
11339-6004               & $Spitzer$ 	SL, SH, LH		   &  11 micron source \\ 
11353-6037              &  $Spitzer$ 	SL, SH, LH		   &   PAH+Sil \\
11387-6113               & $Spitzer$ SL, SH, LH			   &   PAH+Sil \\
12145-5834              & 	$Spitzer$ SL, LL		   &    PAH \\
13245-5036               & 	$Spitzer$ SL, LL		   & Not classified \\
13313-5838               & 	$Spitzer$ SL, LL		   &    PAH \\
13500-6106               & 	$Spitzer$ SL, SH, LH		   &   Sil absorption \\
13529-5934               &	$Spitzer$ SL, SH, LH		   &   Not classified \\
14325-6428                &	$Spitzer$ SL, SH, LH		   &  11 micron source \\
14341-6211               & 	$Spitzer$ SL, SH, LH		   &  11 micron source \\
14346-5952               &	$Spitzer$ SH, LH		   &   Not classified \\
14429-4539               &	$Spitzer$ SL, SH, LH		   &   PAH \\
14482-5725              & $Spitzer$ SL, LL			   &   PAH+Sil \\
15482-5741              & $Spitzer$ SL, LL			   &   PAH \\
16206-5956             & $Spitzer$ SL, SH, LH			   &   Sil emission \\
16494-3930             & $Spitzer$ SH, LH			   &  11 micron source \\
16559-2957               & $Spitzer$ SL, SH, LH			   &  Not classified\\
17009-4154             & 	$Spitzer$ SH		   &   PAH \\
17074-1845               &	$Spitzer$ SL, LL		   & Sil emission\\
17088-4221                  & 	$Spitzer$ SH		   & Not classified\\
17130-4029               & $Spitzer$ SL, SH, LH			   &   PAH \\
17168-3736               &	$Spitzer$ SH, LH		   &  Sil absorption \\
17195-2710               & 	$Spitzer$ SL, SH, LH		   &  Sil absorption \\
17203-1534                  & 	$Spitzer$ SL, LL		   &   Sil emission  \\
17234-4008               & 	$Spitzer$ SL, SH, LH		   &  Sil absorption \\
17253-2831               & 	$Spitzer$ SL, LL, SH, LH		   &  Sil absorption \\
17287-3443              & 	$Spitzer$ SL, SH		   & Transition object with peculiar dust\\
17311-4924                 & 	$Spitzer$ SL, LL		   & Transition object with peculiar dust\\
17317-2743                 &	$Spitzer$ SL, SH, LH		   & Sil emission  \\
17359-2902               & 	$Spitzer$ SL, SH, LH		   &  11 micron source\\
17364-1238              &	$Spitzer$ SL, LL		   &  Sil emission \\
17376-2040              & 	$Spitzer$ SL, LL		   &   Not classified \\
17381-1616               & 	$Spitzer$ SL, LL		   &   Sil emission  \\ 
17423-1755               &	$Spitzer$ SL, LL, $ISO$		   &   featureless \\
17488-1741               &  $Spitzer$ SL, LL			   & Not classified\\
17542-0603               & $Spitzer$ SL, LL			   &  PAH+Sil \\
17580-3111               & 	$Spitzer$ SL, SH, LH		   &  Not classified \\
18023-3409                & 	$Spitzer$ SL, LL		   & Sil emission \\
18062+2410                 &  $Spitzer$ SL, LL                       &  Sil emission \\
18246-1032              &   $Spitzer$ SL, SH, LH                       &   Sil absorption \\
18533+0523               & 	$Spitzer$ SL, SH, LH		   &   PAH \\
19024+0044               & 	$Spitzer$ SL, SH, LH		   &   PAH + sil \\
19157-0247               & 	$Spitzer$ SL, LL, SH, LH		   &   Sil emission\\
19200+3457               & 	$Spitzer$ SL, LL		   &   PAH \\
19306+1407               & 	$Spitzer$ SL, LL,	SH	   &   PAH+Sil\\
19454+2920               & 	$Spitzer$ SH		   &   featureless\\ 
19477+2401               & 	$Spitzer$ SH, LH		   & Not classified\\
19590-1249               & 	$Spitzer$ SL, LL		   & Sil emission \\
20259+4206               & 	$Spitzer$ SH, LH		   & featureless\\
20462+3416               & 	$Spitzer$ SL, LL		   &  PAH+Sil \\
20572+4919               & 	$Spitzer$ SL, LL		   &  Sil emission \\
21289+5815               & 	$Spitzer$ SL, LL		   &  PAH+Sil\\
21546+4721              & 	$Spitzer$ SL, LL		   &  PAH \\
22023+5249               & 	$Spitzer$ SL, LL		   &  PAH+Sil \\
22036+5306              & 	$Spitzer$ SH, LH		   &  Sil absorption \\
22223+4327               &  $Spitzer$ SL, LL, SH, LH			   &  PAH \\
F22327-1731              & 	$Spitzer$ SL, SH, LH		   &  Sil emission\\ 
23304+6147               & 	$Spitzer$ SH, LH, $ISO$		   &   PAH\\
\hline
\hline 
\end{longtable}
\end{center}

\clearpage
\begin{table*}
\begin{center}
\caption{Equivalent widths (in $\mu$m) and peak wavelengths (in $\mu$m) of PAH features in the sample PAGB candidates}
\label{pagb2}
\begin{tabular}{l c c c c c c}
\noalign{\medskip}
\hline\hline
Name	&	\multicolumn{3}{c}{Equivalent width}& \multicolumn{3}{c}{Peak wavelength} \\
\cline{2-7}
& 6.2 $\mu$m        &   7.7 $\mu$m           &  11.2 $\mu$m   & 6.2 $\mu$m    & 7.7 $\mu$m &  11.2 $\mu$m \\
\hline
01005+7910   & 0.380$\pm$0.034 &  0.379$\pm$0.020  &   0.058$\pm$0.005  &  6.225$\pm$0.003   &  7.555$\pm$0.006  &  11.241$\pm$0.003 \\
04296+3429   & 0.158$\pm$0.007 &  0.183$\pm$0.025  &   0.092$\pm$0.008  &  6.262$\pm$0.002   &  7.829$\pm$0.068  &  11.369$\pm$0.013 \\
05113+1347   &       -         &	    -	   &   0.040$\pm$0.005  &	     -       &    - 	         &  11.355$\pm$0.004 \\
05341+0852   &       -         &	    -	   &   0.082$\pm$0.006  &	     -       &    - 	         &  11.352$\pm$0.003 \\
06530-0213   &       -         &	    -	   &   0.132$\pm$0.027  &	     -       &    - 	         &  11.357$\pm$0.007 \\
07134+1005   &       -         &	    -	   &   0.194$\pm$0.017  &	     -       &    - 	         &  11.346$\pm$0.004 \\
08335-4026   & 0.078$\pm$0.006 &  0.108$\pm$0.020  &   0.031$\pm$0.006  &  6.239$\pm$0.003   &  7.703$\pm$0.030  &  11.243$\pm$0.002 \\
11353-6037   & 0.503$\pm$0.019 &  0.514$\pm$0.028  &   0.074$\pm$0.005  &  6.252$\pm$0.009   &  7.773$\pm$0.004  &  11.258$\pm$0.003 \\
11387-6113   & 0.056$\pm$0.006 &  0.050$\pm$0.005  &		-	&  6.307$\pm$0.003   &  7.893$\pm$0.021  &	   -		       \\
12145-5834   & 0.438$\pm$0.016 &  0.368$\pm$0.029  &   0.112$\pm$0.009  &  6.235$\pm$0.002   &  7.597$\pm$0.006  &  11.255$\pm$0.007 \\
13313-5838   & 0.019$\pm$0.006 &	-	   &	      - 	&  6.285$\pm$0.006   &         -	 &	   -		     \\
14429-4539   & 0.216$\pm$0.030 &  0.356$\pm$0.023  &   0.100$\pm$0.019  &  6.267$\pm$0.006   &  7.938$\pm$0.007  &  11.311$\pm$0.006 \\
14482-5725   & 0.041$\pm$0.008 &  0.105$\pm$0.020  &   0.048$\pm$0.006  &  6.298$\pm$0.003   &  8.164$\pm$0.012  &  11.238$\pm$0.011 \\
15482-5741   & 0.114$\pm$0.025 &  0.266$\pm$0.022  &   0.048$\pm$0.005  &  6.262$\pm$0.010   &  7.968$\pm$0.009  &  11.297$\pm$0.009 \\
17009-4154   &      -          &      - 	   &   0.049$\pm$0.006  &	  -	     &        - 	 &  11.253$\pm$0.003 \\
17130-4029   & 0.051$\pm$0.007 &  0.174$\pm$0.014  &	       -	&  6.275$\pm$0.009   &  8.065$\pm$0.013  &	    -	   \\
17542-0603   & 0.011$\pm$0.007 &  0.013$\pm$0.005  &   0.057$\pm$0.006  &  6.312$\pm$0.002   &  7.899$\pm$0.002  &  11.305$\pm$0.011  \\
18533+0523   & 0.072$\pm$0.009 &  0.562$\pm$0.007  &   0.011$\pm$0.004  &  6.282$\pm$0.002   &  7.991$\pm$0.004  &  11.260$\pm$0.011	 \\
19024+0044   & 0.090$\pm$0.004 &  0.021$\pm$0.004  &	      - 	&  6.265$\pm$0.002   &  7.740$\pm$0.006  &	  -	    \\
19200+3457   & 0.441$\pm$0.009 &  0.241$\pm$0.041  &   0.212$\pm$0.040  &  6.232$\pm$0.002   &  7.590$\pm$0.014  &  11.268$\pm$0.004 \\
19306+1407   & 0.339$\pm$0.005 &  0.251$\pm$0.021  &   0.006$\pm$0.003  &  6.239$\pm$0.007   &  7.820$\pm$0.008  &  11.242$\pm$0.011 \\
20462+3416   & 0.215$\pm$0.025 &  0.343$\pm$0.007  &   0.012$\pm$0.064  &  6.233$\pm$0.007   &  7.692$\pm$0.018  &  11.254$\pm$0.018 \\
21289+5815   & 0.100$\pm$0.029 &  0.069$\pm$0.005  &   0.016$\pm$0.005  &  6.241$\pm$0.007   &  7.832$\pm$0.017  &  11.238$\pm$0.014 \\
21546+4721   & 0.170$\pm$0.009 &  0.117$\pm$0.034  &   0.007$\pm$0.002  &  6.226$\pm$0.002   &  7.592$\pm$0.014  &  11.226$\pm$0.002 \\
22023+5249   & 0.304$\pm$0.038 &  0.373$\pm$0.058  &   0.025$\pm$0.005  &  6.220$\pm$0.002   &  7.743$\pm$0.009  &  11.218$\pm$0.006 \\
22223+4327   & 0.025$\pm$0.004 &  0.099$\pm$0.009  &   0.032$\pm$0.006  &  6.267$\pm$0.004   &  7.880$\pm$0.028  &  11.287$\pm$0.004 \\
23304+6147   &     -           &     -  	   &   0.089$\pm$0.004  &	 -	     &        - 	 &  11.360$\pm$0.008 \\
\hline
\noalign{\smallskip}
\end{tabular}
\end{center}
\end{table*}

\clearpage
\begin{table}
\centering
\caption[Circumstellar parameters derived using DUSTY with single shell for a sample of PAGB candidates]{Parameters derived from SED modeling of
 PAGB candidates using DUSTY; the spectral types are from SIMBAD 
and \citet{Suarez06} and both the total and dust mass-loss rates are in the unit of M$_{\odot}$yr$^{-1}$ } 
\label{pagb_single}
\begin{tabular}{l c c c c c c c c c}
\hline\hline
Object &  Sp.Type  & T$_{*}$ & T$_{d}$ & $\tau_{0.55}$ & Grain type  &  r$_{1}$(m) &  r$_{1}$/r$_{c}$  &  \.{M}  & \.{M$_{d}$} \\
\hline
05089+0459   &  M3I	& 3500  &  200 &  6.0  &  AmC(1.0)	    &	2.1E+14 &   1.1E+03  &  5.8E-05                 	    &    4.4E-09\\
06034+1354   &  --	& 5000  &  300 &  1.5  &  AmC(0.9)+Sil(0.1)  &	7.6E+13 &   8.2E+02  &  1.9E-05 			 &    2.2E-09\\
06556+1623   & Bpe        & 15000 & 240 & 0.3 & AmC(0.6)+Gr(0.4) & 2.1E+14 & 2.1E+04 & 1.5E-05 & 6.1E-10\\
08005-2356   &  F5e	& 6650  &  440 &  5.0  &  Gr(0.9)+SiC(0.1)	    &	5.4E+13  &   1.0E+03  &  3.1E-05 			 &    8.1E-10\\
10211-5922 & -- & 5000 & 250 & 0.3 & Gr(0.9)+Sil(0.1) & 1.1E+14 & 1.2E+03 & 7.5E-06 & 1.4E-10\\
11201-6545  & A3Ie      &   6000  &    150  &   5.0   &  Sil(0.5)+AmC(0.5) 	     &    3.5E+14   &   5.5E+03    &    8.7E-05  	 &    2.0E-09\\
11339-6004   &  --	& 14000 &  190 &  8.0  &  AmC(0.9)+SiC(0.1)	    &	4.0E+14 &   3.4E+04  &  1.0E-04 			 &    7.6E-08\\
13245-5036   &  A7Ie	& 7650  &  200 &  7.0  &  AmC(1.0)	    &   2.9E+14 &   7.5E+04  &  8.2E-05 			 &    1.5E-10\\
13500-6106   &  --	& 11000 &  280 &  20.0 &  AmC(0.1)+Gr(0.1)+Sil(0.8)  &	8.2E+13 &   4.2E+03  &  1.0E-04 		 &    2.9E-08\\
13529-5934  &  --       &   12000 &    200  &	13.0  &   AmC(0.2)+Gr(0.8)   &    3.6E+14   &   2.2E+04    &    1.2E-04  	 &    1.8E-08\\
14325-6428  &  F5I      &   6640  &    160  &   2.5   &   AmC(0.9)+SiC(0.1)   &    5.0E+14   &   9.6E+03    &    6.6E-05  	 &    1.6E-09\\
14341-6211   &  --	& 6000  &  195 &  10.0 &  AmC(0.9)+SiC(0.1)	    &	2.9E+14 &   4.5E+03  &  9.5E-05 			 &    1.9E-08\\
14346-5952  & -- & 12000 & 500 & 10.0 & Gr(0.9)+AmC(0.1) & 5.0E+13 & 3.1E+03 & 4.0E-05& 5.5E-10\\
16206-5956   &  A3Iab:e & 8000  &  105 &  0.7  &  AmC(0.3)+Sil(0.7) &	7.6E+14 &   2.1E+04  &  5.1E-05 			 &    1.9E-10\\
16494-3930  &  G2I      &   5160  &    200  &   8.0   &   AmC(0.8)+Gr(0.1)+SiC(0.1)   &    2.5E+14   &   2.9E+03    &    7.9E-05  &    8.2E-08\\
16559-2957   &  F5Iab:e & 6650  &  220 &  4.0  &  AmC(0.7)+Sil(0.3) &	1.6E+14 &   3.1E+03  &  5.1E-05 			 &    1.0E-08\\
17074-1845   &  B5Ibe	& 11000 &  140 &  3.0  &  Sil(1.0)	    &	3.3E+14 &   1.7E+04  &  1.0E-04 			 &    2.1E-09\\
17088-4221  &  --       &   14000 &    250  &	15.0  &   AmC(0.1)+Gr(0.9)   &    2.4E+14   &   2.1E+04    &    1.0E-04  	 &    6.1E-08\\
17168-3736  &  --       &   14000 &    190  &	12.0  &   Gr(0.9)+Sil(0.1)	     &    3.9E+14   &   3.3E+04    &    1.2E-04  	 &    9.4E-08\\
17195-2710   &  --	& 8000  &  370 &  9.0  &  Gr(0.9)+Sil(0.1)	    &   8.1E+13 &   2.2E+03  &   	5.0E-05		 &    2.6E-09\\
17203-1534   &  B1IIIpe & 10000 &  130 &  2.5  &  Gr(0.3)+Sil(0.7)  &	5.0E+14 &   2.1E+04  &  9.0E-05 			 &    1.4E-09\\
17234-4008   &  --	& 14000 &  175 &  10.0 &  Gr(0.9)+Sil(0.1)	    &	4.6E+14 &   3.9E+04  &  1.2E-04 			 &    1.7E-08\\
17253-2831   &  --	& 5000  &  150 &  4.0  &  AmC(0.7)+Gr(0.2)+Sil(0.1)  &	4.5E+14 &   4.9E+03  &  7.6E-05 			 &    1.3E-09\\
17287-3443   &  --	& 18000 &  150 &  20.0 &  AmC(0.3)+Gr(0.7)  &	7.4E+14 &   1.0E+05  &  2.0E-04 			 &    1.9E-08\\
17311-4924   &  B1Iae	& 15000 &  200 &  3.0  &  Gr(0.4)+Sil(0.6)  &	2.3E+14 &   2.9E+04  &  6.7E-05 			 &    2.4E-10\\
17317-2743   &  F5I	& 6640  &  140 &  7.5  &  Gr(0.9)+Sil(0.1)           &   5.2E+14 &   9.9E+03  &  1.1E-04 			 &    4.1E-10\\
17359-2902  &  --       &   8000  &    230  &	15.0  &   Gr(0.9)+SiC(0.1)	     &    2.4E+14   &   6.6E+03    &    1.0E-04  	 &    5.3E-09\\
17364-1238  &  --       &   8000  &    130  &	1.0   &   AmC(0.1)+Sil(0.9)  &    3.2E+14   &   8.9E+03    &    4.6E-05  	 &    3.4E-10\\
17376-2040  &  F6I      &   6460  &    280  &   4.0   &   AmC(0.5)+Gr(0.5)   &    1.2E+14   &   2.2E+03    &    4.1E-05  	 &    9.1E-10\\
17381-1616   &  B1Ibe	& 20700 &  400 &  1.0  &  AmC(0.9)+Sil(0.1)	    &	5.7E+13 &   1.0E+04  &  1.7E-05 			 &    5.1E-09\\
17423-1755   &  Be	& 13000 &  170 &  5.0  &  Gr(0.8)+Sil(0.2)  &	4.0E+14 &   2.6E+04  &  8.6E-05 			 &    7.7E-10\\
17488-1741  &  F7I      &   6280  &    260  &   6.5   &   Gr(1.0)	     &    1.6E+14   &   2.8E+03    &    6.0E-05  	 &    1.6E-10\\
17580-3111 & -- & 5000 & 290 & 15.0 & Gr(1.0) & 1.1E+14 & 1.2E+03 & 7.3E-05 & 1.2E-08\\
18023-3409   &  B2IIIe  & 10000 &  175 &  1.0  &  AmC(0.3)+Sil(0.7) &	2.2E+14 &   9.6E+03  &  3.7E-05 			 &    3.0E-09\\
18062+2410   &  B1IIIpe & 18000 &  200 &  1.0  &  Sil(1.0)	    &	1.9E+14 &   2.6E+04  &  4.5E-05 			 &    4.0E-08\\
18246-1032  &  --       &   20000 &    130  &	20.0  &   Gr(0.8)+Sil(0.2)   &    8.5E+14   &   1.3E+05    &    2.2E-04  	 &    2.6E-08\\
19157-0247   &  B1III	& 15000 &  800 &  2.0  &  Gr(0.8)+Sil(0.2)  &	1.5E+13 &   1.5E+03  &  1.2E-05 			 &    1.5E-09\\
19454+2920   &  --	& 10000 &  180 &  7.0  &  AmC(1.0)	    &	4.2E+14 &   1.8E+04  &  9.7E-05 	  		 &    4.1E-08\\
19477+2401   &  F4I	& 6820  &  200 &  10.0 &  AmC(0.8)+Gr(0.2)  &   2.8E+14 &   5.7E+03  &  9.7E-05 			 &    2.7E-10\\
19590-1249   &  B1Ibe	& 20000 &  125 &  0.1  &  Gr(0.3)+Sil(0.7)  &	7.1E+14 &   1.2E+05  &  1.8E-05 			 &    1.0E-09\\
20259+4206  &  F3I      &   6990  &    200  &   8.0   &   AmC(1.0)	     &    2.9E+14   &   6.1E+03    &    8.5E-05  	 &    2.3E-08\\
20572+4919   &  F3Ie	& 7700  &  265 &  2.0  &  Gr(0.85)+Sil(0.15)&   1.2E+14 &   2.9E+03  &  2.9E-05 			 &    8.3E-10\\
22036+5306   &  --	& 8000  &  140 &  8.0  &  AmC(0.7)+Gr(0.2)+Sil(0.1)  &   6.7E+14 &   1.8E+04  &  1.3E-04    			 &    1.3E-09\\
F22327-1731  &  A0III	& 7600  &  600 &  0.35 &  Gr(0.85)+Sil(0.15)&	2.2E+13 &   5.5E+02  &  4.6E-06 	   		 &    7.5E-10\\
\hline
\end{tabular}
\end{table}
\clearpage

\begin{landscape}
\begin{table}
\centering
\caption[Circumstellar parameters derived using DUSTY with two shells for a sample of PAGB candidates]{Parameters derived with two shells in the SED modeling using DUSTY for the PAGB candidates; the spectral types are from SIMBAD and both the total mass-loss rates are in the unit of M$_{\odot}$yr$^{-1}$} 
\label{ppne_2shells}
\begin{tabular}{l c c c c c c c c c c r}
\toprule
Objects & Sp. Type & $T_{*}$ & \multicolumn{4}{c}{Sphere1} & \multicolumn{4}{c}{Sphere2} &   \.{M}  \\
\cline{4-11}
& & & PAH type & $T_{d}$ & $r_{1}$ (m) & $r_{1}$/$r_{c}$ & Grain type & $T_{d}$ & $r_{1}$ (m) & $r_{1}$/$r_{c}$ \\
\midrule
01005+7910 &	B2Iab:e	& 15000 &  ionised &	   700  &   3.1E+13    &    3.0E+03	&     Gr(1.0)		  &    250  &	2.5E+14 &	 2.4E+04 &4.0E-05\\
04296+3429 &	G0Ia	& 5510  &  ionised &	   600  &   1.9E+13	&    2.5E+02	&     AmC(0.9)+Gr(0.1)    &    200  &	2.6E+14 &	 3.4E+03 &6.6E-05\\
05113+1347 &	G8Ia	& 4900  &  neutral &	   250  &   1.5E+14	&    1.6E+03	&     AmC(1.0)  	  &    200  &	2.4E+14 &	 2.5E+03 &4.7E-05\\
05341+0852 &	F4Iab	& 6820  &  neutral &	   500  &   4.2E+13	&    8.4E+02	&     Gr(1.0)		  &    350  &	9.6E+13 &	 1.9E+03 &3.4E-05\\
06530-0213 &	F0Iab:	& 7300  &  neutral &	   250  &   2.4E+14	&    5.6E+03	&     Gr(1.0)		  &    100  &	1.2E+15 &	 2.8E+04 &5.0E-05\\
07134+1005 &	F5Iab:	& 6650  &  neutral &	   600  &   2.5E+13	&    4.7E+02	&     AmC(1.0)  	  &    150  &	6.1E+14 &	 1.1E+04 &3.4E-05\\
08335-4026     & B8e   & 11500   & ionised	   & 700     & 2.7E+13        & 1.5E+03	 & Gr(0.95)+Sil(0.05)	   & 100     & 1.3E+15        & 7.9E+04  &1.2E-05\\
11353-6037     & B5Ie  & 13400   & ionised	   & 500     & 7.3E+13        & 5.6E+03	 & Gr(1.0)		   & 220     & 3.0E+14        & 2.3E+04  &6.4E-05\\
11387-6113   & A3I  & 8890 & neutral & 190 & 5.4E+14 & 1.85E+04 & Gr(0.8)+Sil(0.2) & 150 & 4.7E+14 & 1.6E+04                                             &7.2E-05\\
12145-5834     & -     & 10000   & ionised	   & 500     & 5.9E+13        & 2.5E+03	 & Gr(1.0)		   & 100     & 1.3E+15        & 5.7E+04  &7.4E-05\\
13313-5838 &	K1III	& 4600  &  neutral &	   400  &   4.8E+13	&    4.4E+02	&     Gr(0.2)+AmC(0.8)    &    200  &	2.3E+14 &	 2.1E+03 &3.1E-05\\
14429-4539 &	F4I	& 6820  &  ionised &	   400  &   7.0E+13	&    1.4E+03	&     Gr(0.6)+AmC(0.4)    &    320  &	9.7E+13 &	 1.9E+03 &4.1E-05\\
14482-5725     & A2I   & 9080	 & neutral	   & 400     & 4.2E+13        & 2.9E+02	 & Gr(1.0)		   & 300     & 9.1E+13        & 6.2E+02  &3.6E-05\\
15482-5741     & F7I   & 6100	 & ionised	   & 400     & 6.1E+13        & 9.9E+02	 & AmC(1.0)		   & 150     & 5.9E+14        & 9.4E+03  &8.3E-05\\
17009-4154 &	-	& 4500  &  neutral &	   250  &   1.4E+14	&    1.2E+03	&     Gr(1.0)		  &    100  &	8.9E+14 &	 7.8E+03 &8.7E-05\\
17130-4029 & - & 14000 & ionised & 1300 & 5.7E+12 & 4.8E+02 & Gr(0.2)+AmC(0.8) & 120 & 1.2E+15 & 1.0E+05                                                 &6.5E-05\\
17542-0603 &	Ge	& 5500  &  ionised &	   1000 &   5.0E+12	&    6.5E+01	&     Gr(1.0)		  &    500  &	3.7E+13 &	 4.8E+02 &1.6E-05\\
18533+0523 &	-	& 5000  &  ionised &	   400  &   5.0E+13	&    5.4E+02	&     Gr(0.7)+AmC(0.3)    &    100  &	1.0E+15 &	 1.1E+04 &9.4E-05\\
19024+0044 &  - & 10000 & ionised & 1200 & 5.6E+12 & 2.4E+02 & AmC(0.8)+Sil(0.2) & 130 & 8.4E+14  & 3.6E+04                                              &1.1E-04\\
19200+3457 &	B...	& 12000 &  ionised &	   500  &   6.7E+13	&    4.1E+03	&     Gr(1.0)		  &    230  &	2.5E+14 &	 1.5E+04 &1.5E-05\\
19306+1407 &	B0:e	& 15400 &  ionised &	   800  &   2.2E+13	&    2.3E+03	&     Gr(1.0)		  &    100  &	1.6E+15 &	 1.6E+05 &8.9E-05\\
20462+3416 &	B1Iae	& 20800 &  ionised &	   150  &   1.5E+15	&    2.9E+05	&     Gr(0.5)+AmC(0.5)    &    100  &	1.7E+15 &	 3.3E+05 &1.7E-05\\
21289+5815 &	A2Ie	& 7000  &  ionised &	   700  &   1.7E+13	&    3.5E+02	&     Gr(0.7)+Sil(0.3)    &    300  &	8.6E+13 &	 1.8E+03 &3.9E-05\\
21546+4721     & -     & 10000   & ionised	   & 500     & 5.9E+13        & 2.5E+03	 & Gr(1.0)		   & 200     & 3.4E+14        & 1.4E+04  &4.3E-05\\
22023+5249 &	Be	& 12000 &  ionised &	   200  &   6.1E+14	&    3.8E+04	&     Gr(0.2)+AmC(0.8)    &    160  &	5.9E+14 &	 3.6E+04 &8.7E-05\\
22223+4327 &	F9Ia	& 6100  &  ionised &	   700  &   1.4E+13	&    2.3E+02	&     Gr(1.0)		  &    180  &	3.2E+14 &	 5.2E+03 &3.7E-05\\
23304+6147 &	G2Ia	& 5200  &  neutral &	   400  &   5.5E+13	&    6.4E+02	&     Gr(0.5)+AmC(0.5)    &    170  &	3.6E+14 &	 4.1E+03 &4.3E-05\\
\hline
\end{tabular}
\end{table}
\end{landscape}
\clearpage

\appendix
\section{Continuum-subtracted Spitzer IRS spectra of the sample PAGB stars}

The Continuum-subtracted $Spitzer$ IRS spectra in the wavelength region 5 - 38 $\mu$m are shown in the following figures,
arranged in the category-wise (see text for details).

\begin{figure}
\centering
 \includegraphics[width =9.0in,angle=90]{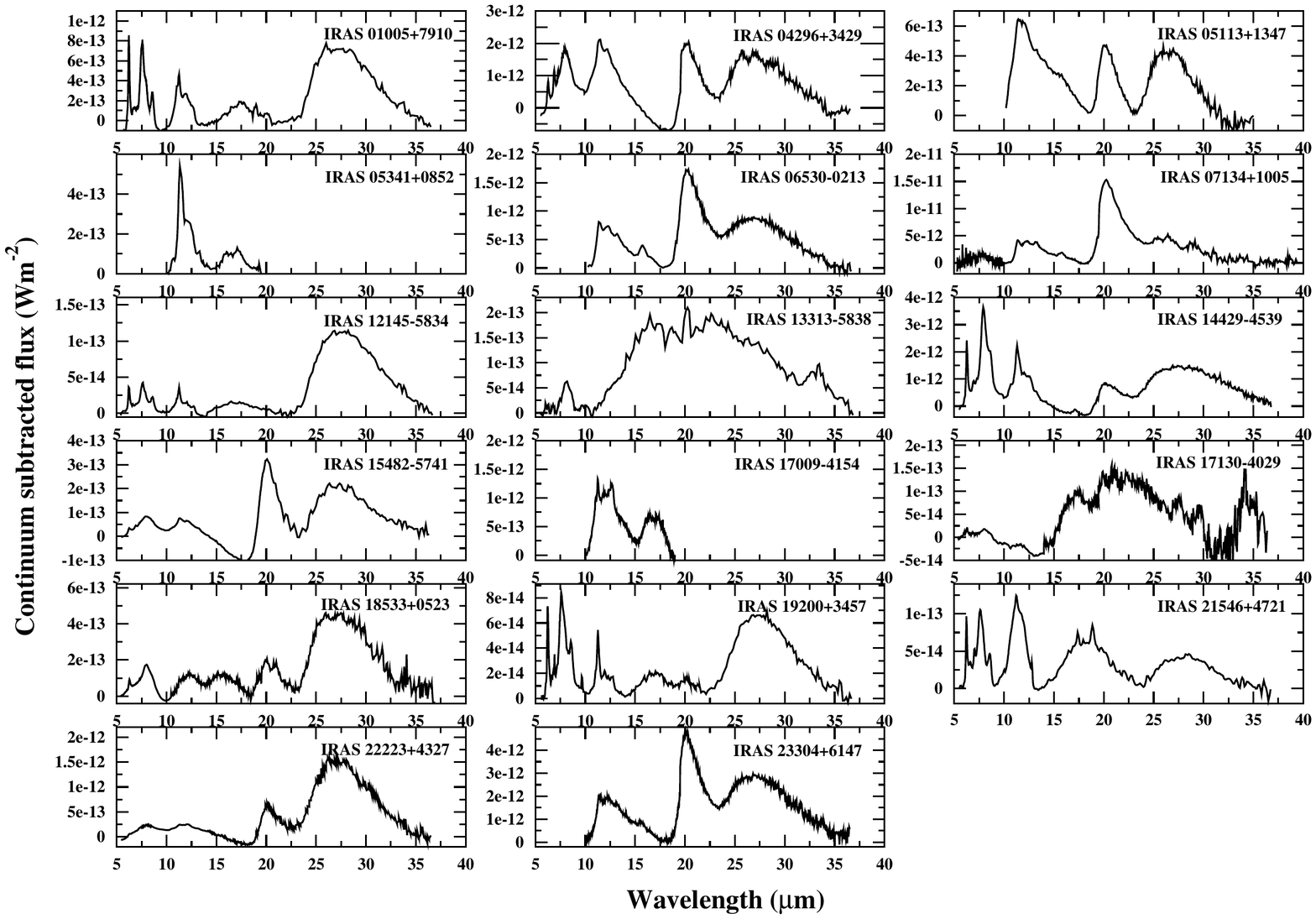}
   \caption{Continuum-subtracted $Spitzer$ IRS spectra for sources with PAH emissions (Classes 1a, 1b, 1c)}
   \label{figa2}%
   \end{figure}
\begin{figure}
  \centering
   \includegraphics[width =9.0in,angle=90]{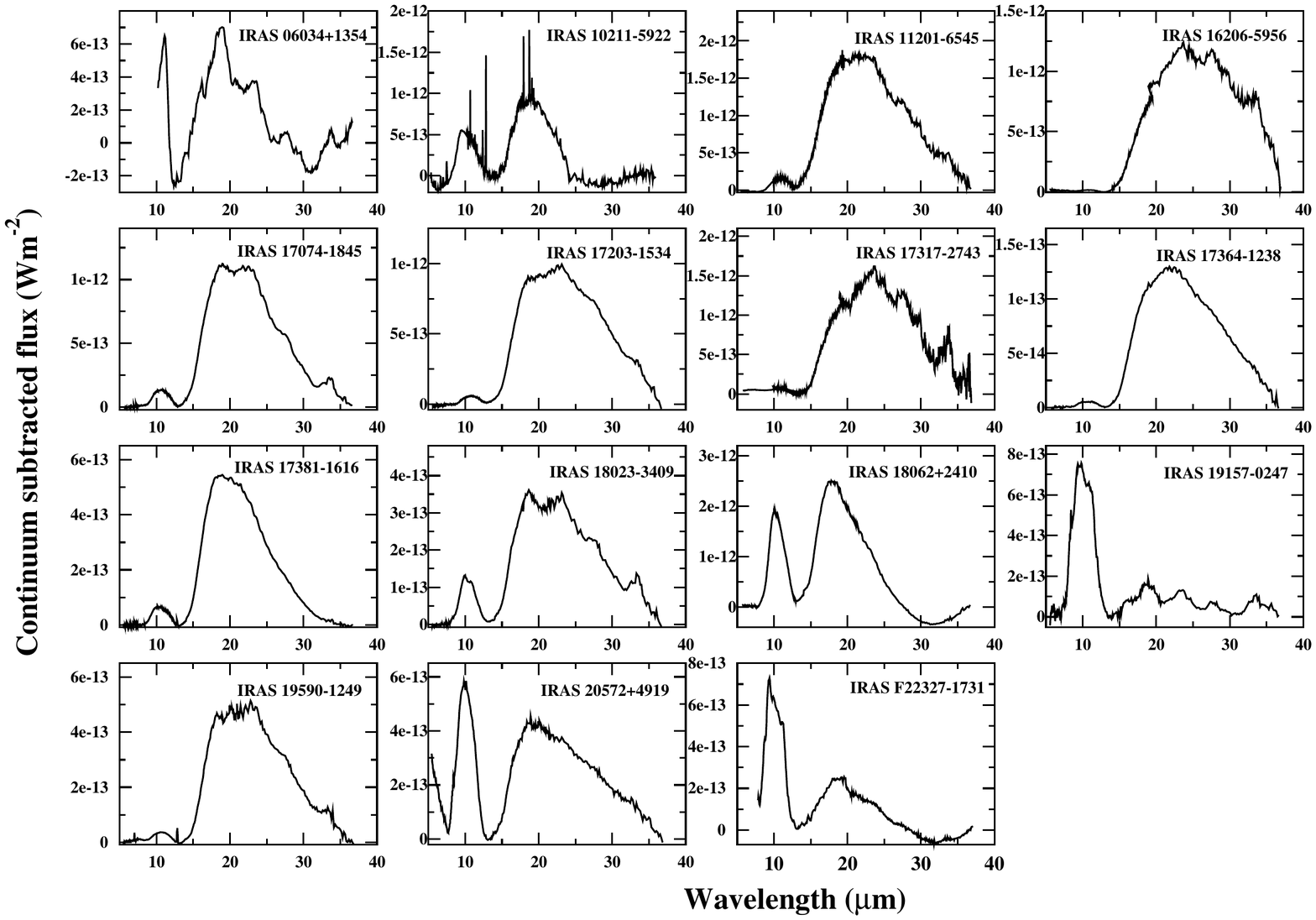}
     \caption{Continuum-subtracted $Spitzer$ IRS spectra for sources with silicate emission (Class 2a)}
     \label{figa8}%
     \end{figure}   

 \begin{figure}
 \centering
  \includegraphics[width =9.0in,angle=90]{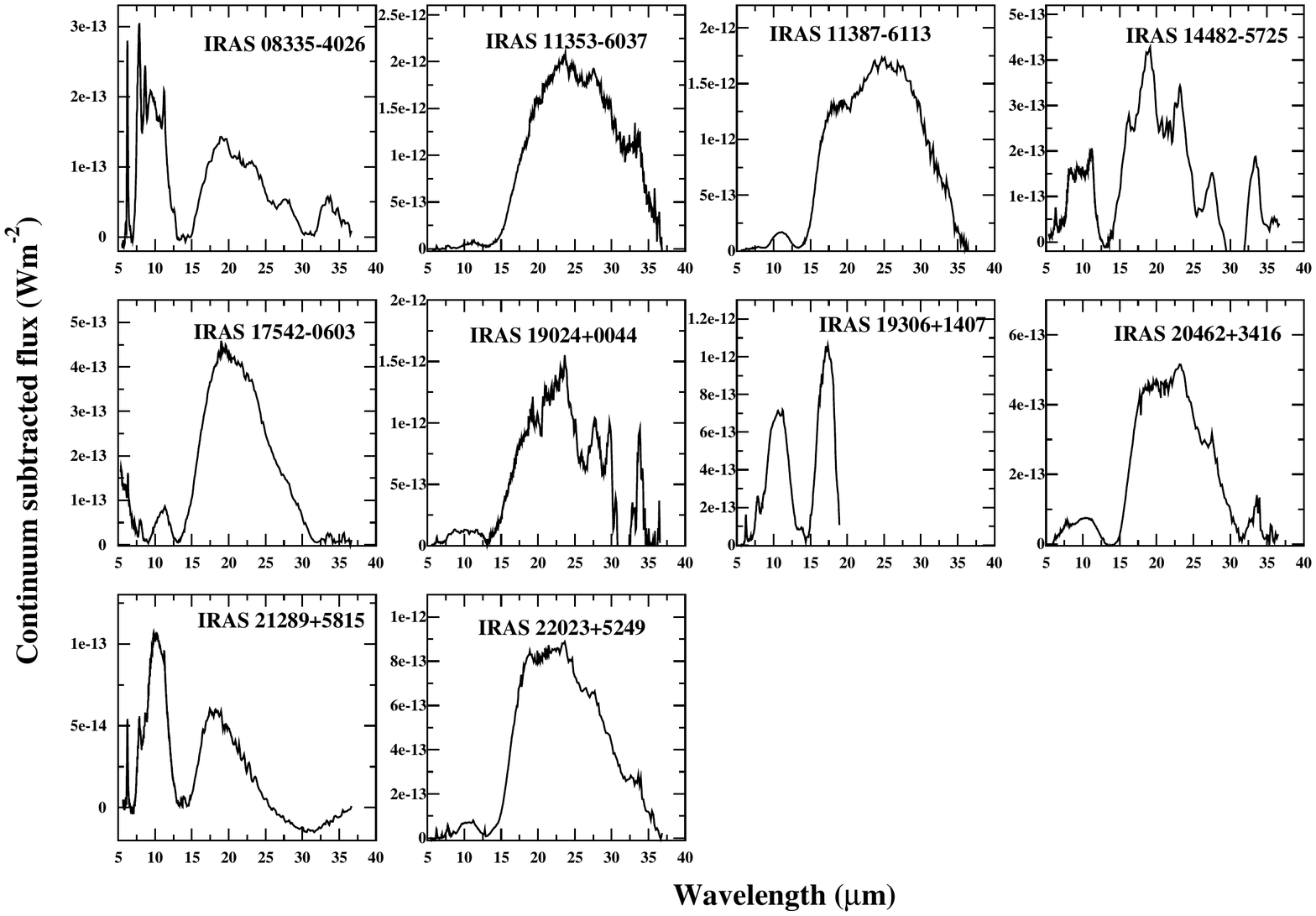}
    \caption{Continuum-subtracted $Spitzer$ IRS spectra for sources with PAH and silicate emissions (Class 2b)}
    \label{figa4}%
    \end{figure}     

 \begin{figure}
  \centering
   \includegraphics[width =9.0in,angle=90]{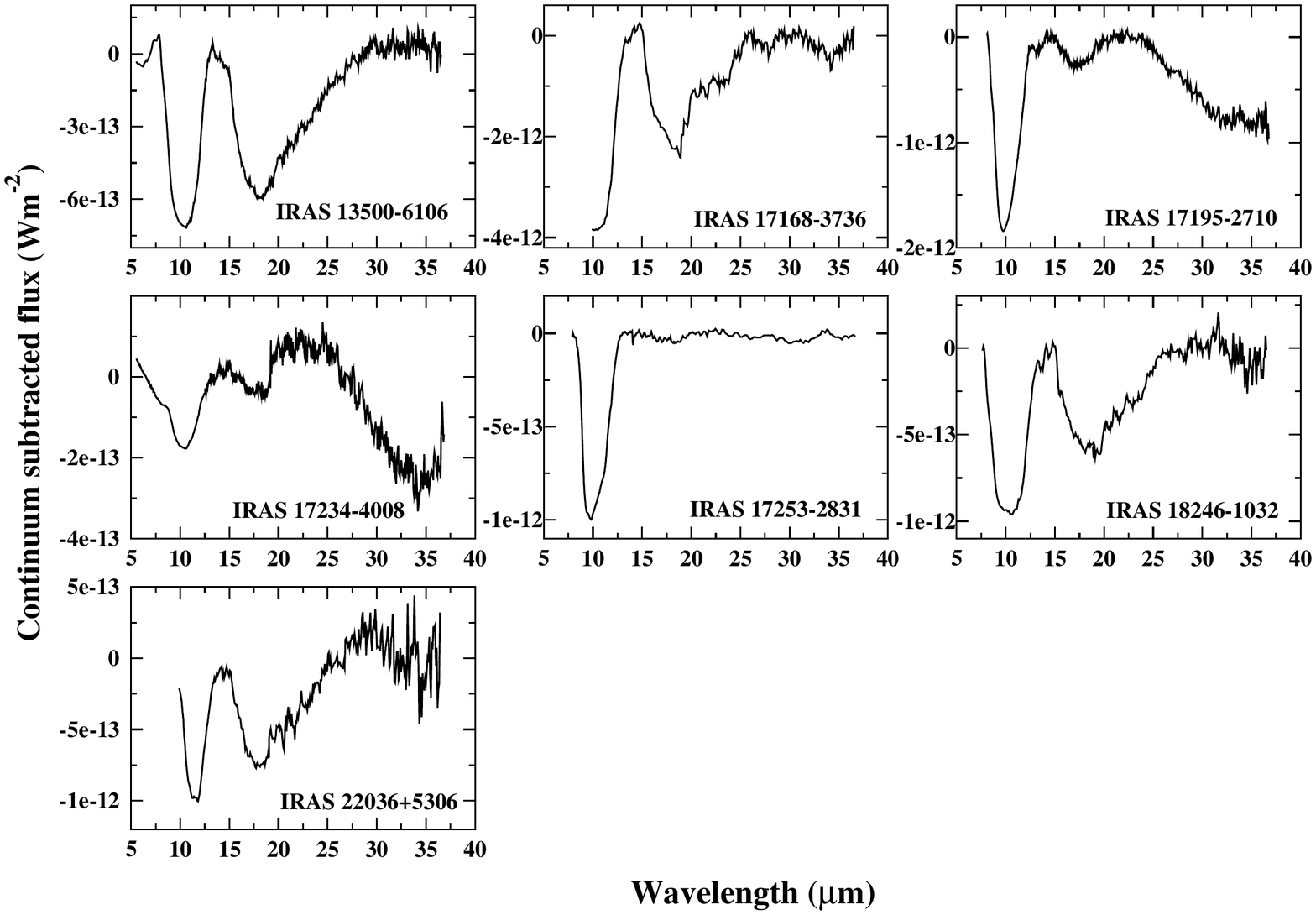}
     \caption{Continuum-subtracted $Spitzer$ IRS spectra for sources with silicate absorption (Class 3)}
     \label{figa6}%
     \end{figure}

  \begin{figure}
       \centering
        \includegraphics[width =9.0in,angle=90]{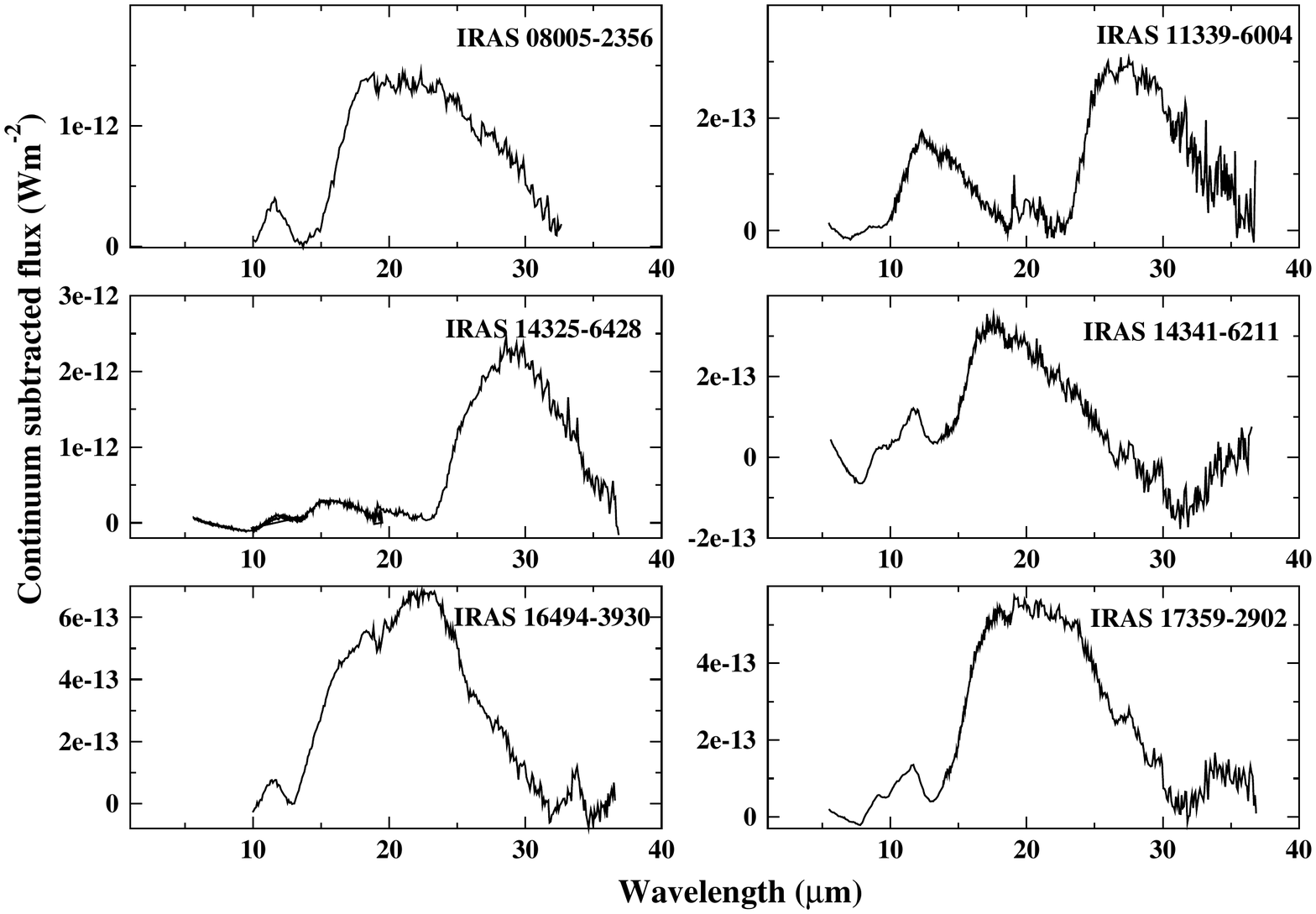}
          \caption{Continuum-subtracted $Spitzer$ IRS spectra for sources with a broad 11 $\mu$m emission (Class 4)}
          \label{figa10}%
  \end{figure}
                         
  \begin{figure}
     \centering
      \includegraphics[width =9.0in,angle=90]{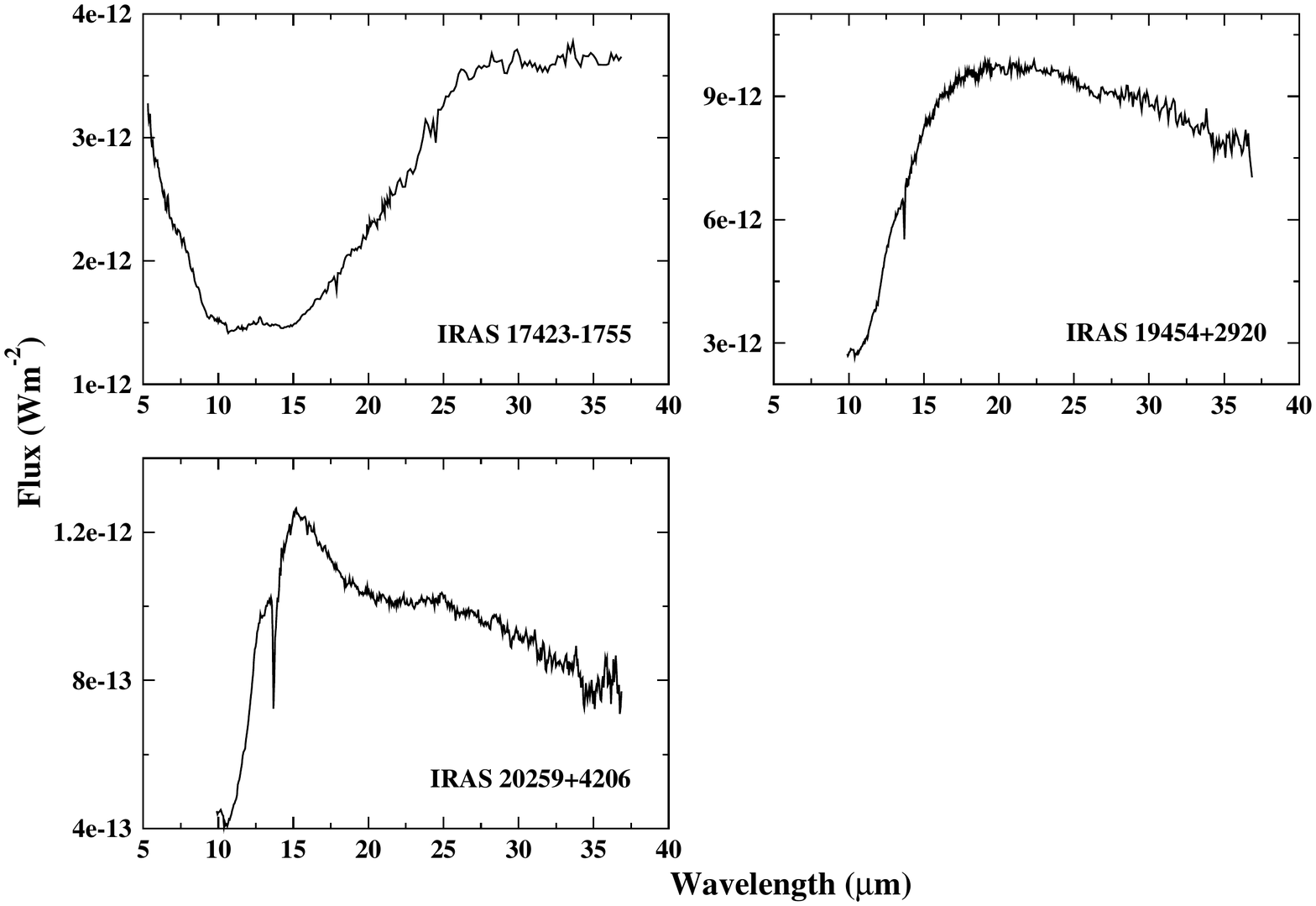}
        \caption{$Spitzer$ IRS spectra for sources with featureless dust continuum (Class 5)}
        \label{figa12}%
        \end{figure} 
\begin{figure}
  \centering
   \includegraphics[width =9.0in,angle=90]{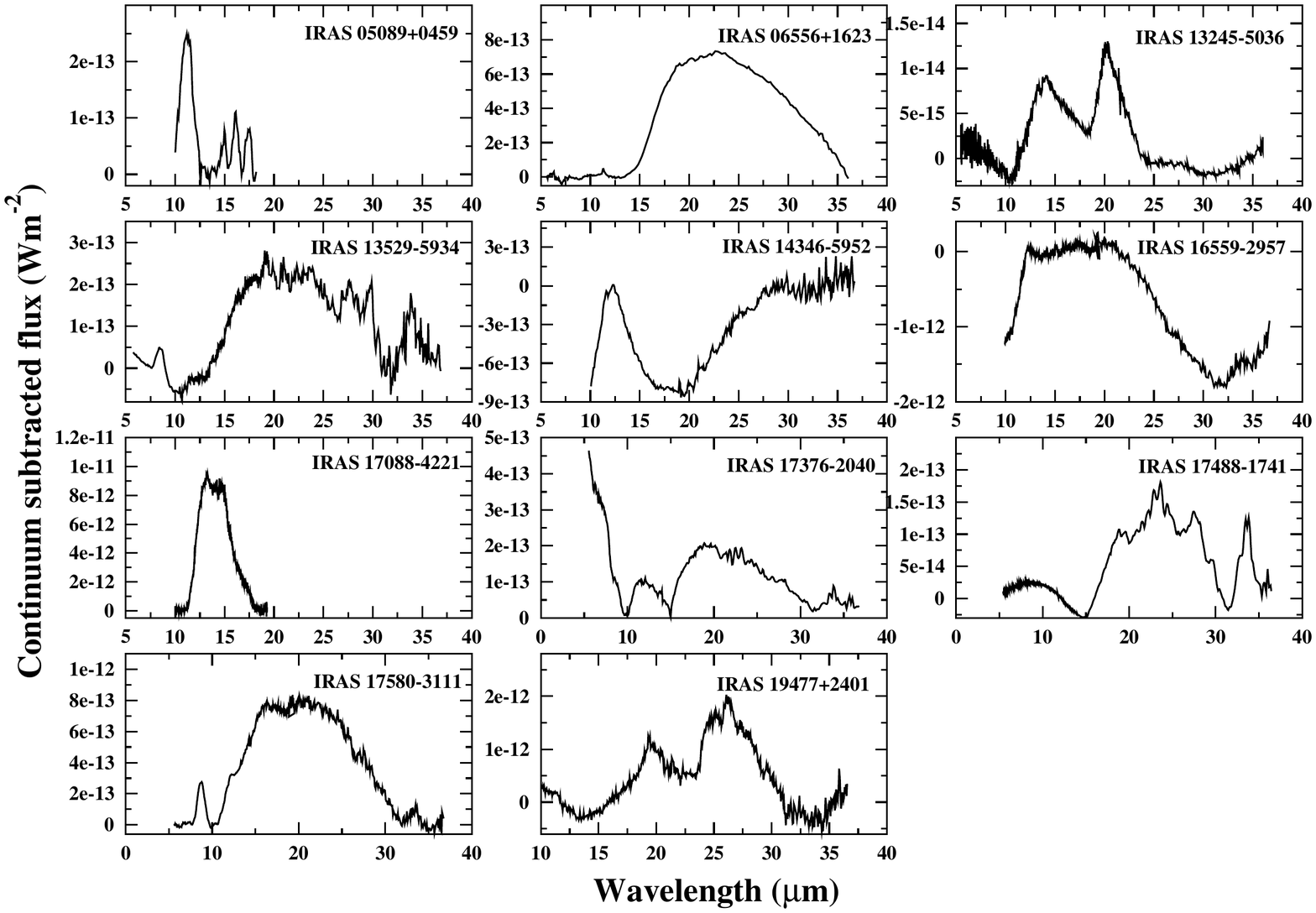}
     \caption{Continuum-subtracted $Spitzer$ IRS spectra for sources that do not fall in the proposed classification scheme}
     \label{figa13}%
     \end{figure}

\end{document}